\documentclass[aps,jcp,reprint,superscriptaddress]{revtex4-2}

\usepackage{graphicx}
\usepackage{eqnarray,amsmath}
\usepackage{dcolumn}
\usepackage{bm}
\usepackage{amssymb}
\usepackage{alphabeta}
\usepackage{comment}
\usepackage[nolist, nohyperlinks]{acronym}
\usepackage[mathlines]{lineno}
\usepackage[dvipsnames]{xcolor}
\usepackage[version=4]{mhchem}
\usepackage{appendix}
\usepackage{hyperref}
\usepackage{cleveref}
\usepackage{xcolor}

\usepackage{array}

\usepackage{algorithmic}
\usepackage{listings}

\usepackage[normalem]{ulem}

\newcommand{\mm}[1]{\textcolor{red}{#1}}

\newcommand{\svs}[1]{\textcolor{orange}{#1}}

\begin{document}%
\title{Advanced Langevin thermostats: Properties, extensions to rheology,\\ and a lean momentum-conserving approach}

\author{Shubham Agarwal}
\affiliation{Dept. of Materials @ Science and Engineering, Saarland University, Germany}
\author{Sergey V. Sukhomlinov}
\affiliation{Dept. of Materials @ Science and Engineering, Saarland University, Germany}
\author{Marc Honecker}
\affiliation{Dept. of Materials @ Science and Engineering, Saarland University, Germany}
\author{Martin H. Müser}
\affiliation{Dept. of Materials @ Science and Engineering, Saarland University, Germany}

\date{\today}

\begin{abstract}
The Langevin equation accounts for unresolved bath degrees of freedom driving the system toward the bath temperature. 
Because of this, numerical solutions of the Langevin equation have a long history. 
Here, we recapitulate, combine, and extend existing Langevin-equation based thermostats, scrutinize their properties and demonstrate their superiority over global kinetic-energy controls.
Our work includes compact, asymptotic-analysis based derivations of stochastic thermostats, including the highly accurate Grønbech-Jensen scheme.
Proposed extensions include a precise, colored and a lean, momentum-conserving thermostat.
\end{abstract}

\date{\today}
\maketitle

\section{Introduction}

Imposing temperature in molecular dynamics (MD) simulations is a critical task~\cite{AllenTildesley2017book,Frenkel2001book}.
It is commonly achieved through so-called thermostats, which are modifications to the equations of motion or of their solution.
%
In most MD simulations, they are meant to establish the canonical distribution in systems consisting of supposedly coupled degrees of freedom (DOFs).
Despite notable exceptions, thermostats can be roughly grouped into two categories: global kinetic energy controls (GKEC)~\cite{Berendsen1984JCP,Nose1984JCP,Hoover1985PRA,Evans1985JCP,Martyna1992JCP} and local thermostats, which act either on the motion of point masses relative to a frame of reference~\cite{Schneider1978PRB,GronbechJensen2019MP} or on the relative motion between nearby DOFs~\cite{Espanol1995EPL,Lowe1999EPL,Soddemann2003PRE}.
Local thermostats generally hinge on the solution of the Langevin-equation, see Sect.~\ref{sec:langevin}, but also on generalizations thereof~\cite{Ceriotti2009PRL}.
Thermostatting schemes developed for the motion of point masses can also be applied to control the (thermal) motion of auxiliary variables, such as the shape of the simulation cell in a constant-stress simulation~\cite{Parrinello1981JAP}, in which case they are called barostats, or charge-transfer variables, allowing Ohmic losses and Nyquist noise to be modeled during an MD simulation~\cite{Muser2024JCP}.

The crux of thermostats is that they should be sufficiently intrusive to ensure ergodicity in long simulation runs.
At the same time, the dynamics should remain as natural as possible when targeting dynamical properties such as viscosity or the diffusion coefficient.
Broadly speaking, GKECs fail to meet the first requirement, despite three milestone papers~\cite{Hoover1985PRA,Martyna1992JCP,Bussi2007JCP}, having the word \textit{canonical} in the title, while Langevin-based solvers, particularly those violating Galilean invariance, fail to meet the second.

In this study, we reiterate and corroborate claims made above while exploring possibilities for identifying satisfactory compromises between the mutually exclusive requirements of applying a thermostat while remaining non-invasive.  
To this end, much attention will be paid to model systems containing two identical or similar sub-systems, or (collective) DOFs, which couple only weakly to the remaining DOFs.  
Long-wavelength vibrations in a homogeneous medium are one example, e.g., $a(t)\cos(qx)$ and $b(t)\sin(qx)$, where $q$ is the smallest wavenumber fitting into a simulation cell, while $a(t)$ and $b(t)$ are time-dependent amplitudes.  
The larger the system, the less these modes couple to the remaining modes, or music could not be transmitted through the ambient atmosphere.
Global GKECs produce dynamics in which $a(t)$ and $b(t)$ remain correlated, causing uncoupled modes to appear rigidly coupled.
This can have detrimental consequences when computing static response functions, e.g., when the two quasi-harmonic variables are the size of a simulation cell in $x$ and $y$ direction, respectively, and elastic-tensor elements are determined from the covariance of the box shape~\cite{Parrinello1982JCP}. 
While Langevin-based thermostats do not produce erroneous static response functions in the long run, they risk to produce inaccurate dynamics.
%
Momentum-conserving schemes, such as that adopted in dissipative-particle dynamics (DPD)~\cite{Espanol1995EPL,Soddemann2003PRE,Vattulainen2002JCP} alleviate that situation
but tend to be computationally heavy.

Given the above discussion, it seems as though a thermostat that is minimally intrusive on the dynamics, while ensuring highly accurate ergodic sampling and being computationally lean appears to be missing. 
In our opinion, ``colored thermostats à la carte'' (CTs) proposed by Ceriotti, Bussi, and Parrinello~\cite{Ceriotti2009PRL,Ceriotti2010JCTC} come closest to this ideal if their time constants are set such that the most-difficult-to-equilibrate long wavelength modes are near-critically damped during equilibration and underdamped during the observation phase of the simulation.
The basic idea of their approach is to pass damping and random noise through a low-pass filter in a way that the fluctuation-dissipation theorem~\cite{Kubo1966RPP} is satisfied.
In the language of rheology, each thermostatted DOF is coupled to its own thermostatted Maxwell element.
This reduces the artificial, short-time erratic atomic motion, which can be particularly undesirable when auxiliary variables, e.g., the wave function coefficients in Car-Parrinello MD~\cite{Car1985PRL}, are treated with an extended Lagrangian scheme. 
Unfortunately, the original CTs do not produce exact expectation values for harmonic references, unlike the Grønbech-Jensen Langevin (GJL) thermostat~\cite{GronbechJensen2019MP}.
However, both approaches can be readily combined, as we demonstrate in this work.

%
%
%

We also propose what we call the best-possible \textit{conventional} Langevin (BPCL) thermostat.
It is slightly less accurate than GJL but also transitions seamlessly to Brownian dynamics in the limit of small damping times $\Delta t$ while allowing the time step to be increased as the thermostat's time constant decreases. 
However, BPCL does not require random numbers to be remembered from one time step to the next.
This proves useful in the design of a lean momentum-conserving Langevin (MCL) thermostat, in which groups of near-by atoms are thermostatted relative to their center-of-mass motion. 
Such a scheme allows dynamics to remain oscillatory at large wavelength, i.e., hydrodynamic interactions are reduced in a controllable fashion but not suppressed as strongly as with local thermostats acting in a reference frame.
%


While this work is not meant to be a review or a text book, we feel that many readers will appreciate derivations of thermostat algorithms that are easier to follow than those those commonly pursued, which rely on stochastic time propagation operators, their break-up, and Wiener processes~\cite{Skeel2002MP, Ricci2003MP,Melchionna2007JCP}.
Instead, we pursue a pragmatic approach, in which the integrations scheme is defined in a first step using arbitrary propagation coefficients, as in Eqs.~\eqref{eq:conventLang}, \eqref{eq:brown_osc}, and \eqref{eq:gj_scheme}.
In a second step, the coefficients are constraint to produce as many exact results as possible, e.g., any integration scheme should produce the correct drift given a constant external force. 
This allows us to motivate the algorithms with just a few simple rather than many complex equations. 
At the same time, we provide some background theory, which, in our opinion should be part of regular text books, but may yet be difficult to find. 
For a more extended background, we refer the reader to excellent works addressing constraints~\cite{Vanden2006CPL,Goga2012JCTC}, stochastic thermostats for multiple-time stepping~\cite{Izaguirre2001JCP}, and non-equilibrium situations~\cite{Ruiz2018EPJE}.

The remainder of this article is organized as follows.
Sect.~\ref{sec:background} provides the necessary background for this article, including the definition of the Langevin equation, the motivation of various Langevin-based integration schemes, a critical assessment of GKEC algorithms, and a hopefully pedagogical discussion of the damped linear chain, which may even help explain sex-based differences in voice pitch.
Results are presented in Sect.~\ref{sec:results}, where we first focus on static observables, such as the first and second moment of the potential energy, and later on dynamical observables, in particular those related to vibrational frequencies.
The model systems are primarily one-dimensional, such as a linear chain of atoms interacting through a nearest-neighbor Lennard-Jones (LJ) potential~\cite{Schwerdtfeger2024JCTC}. 
They allows us to effectively address issues arising due to large linear system sizes as well as to large anharmonicity, since LJ atoms are essentially underway on a head-on collision at each moment of time.
At the same time, highly accurate averages are readily obtained. 
However, the harmonic oscillator, a sinusoidal potential, and three-dimensional bulk liquids will also be explored. 
The main findings are summarized and conclusions are drawn in Sect.~\ref{sec:conclusions}

\section{Background, Methods and Theory}
\label{sec:background}

\subsection{Langevin Equation}
\label{sec:langevin}

Stochastic thermostats approximately solve 
the ordinary differential equation (ODE) but also coupled ODEs of the form
\begin{equation}
\label{eq:langevin_equation}
m \ddot{x}  = - m \dot{x}/\tau + f(x) + \Gamma(t),
\end{equation}
which is also known as Langevin equation.
Here $\tau$ is a relaxation time and
$m$ is the inertia of a point mass subjected to  damping, deterministic, and thermal random forces.
In the most general case, both $m$ and $1/\tau$ can be matrices, which do not even have to be diagonal.
However, their eigenvalues must be non-negative and they must not share eigenvectors with zero eigenvalues.
Since general features of the solution of Eq.~\eqref{eq:langevin_equation}
 do not depend on the dimensionality of the vector $x$, we first focus on a single degree of freedom. 
In this case, first and second moments  of the random force $\Gamma(t)$ satisfy
\begin{align}
\label{eq:random_force}
\left \langle \Gamma(t) \right\rangle & = 0 \\
\left \langle \Gamma(t) \Gamma(t') \right\rangle & = 2k_B T m \delta(t-t')/\tau ,
\end{align}
respectively, where $k_BT$ is the thermal energy.

\subsection{Conventional Langevin Thermostats}

In this work, any Langevin-equation solver producing a set of positions $x_n$ with the integration scheme 
\begin{subequations}
\label{eq:conventLang}
\begin{eqnarray}
\label{eq:veloc_stepC}
v_{n+1/2} & = & c_{vv} v_{n-1/2} + c_{vf} f_n + c_{vg}g_{n} \\
\label{eq:posit_stepC}
x_{n+1} & = & x_n + c_{xv} v_{n+1/2}
\end{eqnarray}
\end{subequations}
will be denoted as conventional Langevin thermostat.
Here, $c_{vv}$, $c_{vf}$, $c_{vg}$, and $c_{xv}$ are propagation coefficients, whose numerical values are supposed to solely depend on $m$, $\tau$, and $k_BT$. 
Their functional dependencies can change from one algorithm to the next.
The $g_n$ are independent standard Gaussian random numbers, unless stated otherwise.

The notation used in Eq.~\eqref{eq:conventLang} is inspired by the way in which Allen and Tildesley~\cite{AllenTildesley2017book} recast the leapfrog Verlet algorithm as two successive mid-point integrations.
It is possible to rewrite the algorithm using positions only, as is done in the Verlet scheme. 
However, the current formulation is seen as advantageous, because both the number of floating point operations and the numerical-roundoff errors are minimal. 
This is why we focus entirely on this variant of a conventional Langevin thermostat. 

The employed convention for the propagation coefficients $c_{\alpha\beta}$ makes unit checks simple, because the unit of a coefficient is that of its first divided by that of its second index.
Thus, an expression of the form $c_{xv} c_{vg} g $ is easily recognized to have the same unit as $x$.
To facilitate calculations conducted later when considering the harmonic oscillator, we will also use 
\begin{equation}
c_{vx} = - c_{vf} k,
\end{equation}
where $k$ is the spring constant. 
Another notation meant to streamline the presentation is that
$\langle a b \rangle$ denotes the expectation value of $a_n b_n$ in thermal equilibrium, where $a$ and $b$ can represent any observable like $x$ or $v$. 
Moreover, upper $\pm$ indices are used to denote covariances, where one observable is taken a half-time step before or after the other:
\begin{equation}
\langle a^\pm b \rangle \equiv \langle a_{n\pm 1/2} b_n \rangle.
\end{equation}
Self covariances are symmetric in time so that
\begin{equation}
\langle v^+ x \rangle = - \langle v^- x \rangle.
\end{equation}

A few general rules can be identified for covariances in conventional Langevin thermostats, which proof useful when studying the properties of specific $c_{\alpha\beta}$ parameterizations.
Squaring Eq.~\ref{eq:veloc_stepC} and taking expectation values 
yields
\begin{eqnarray}
\label{eq:veloc_squared}
(1-c_{vv}^2) \langle v^2 \rangle & = & c_{vg}^2 + c_{vf}^2 \langle f^2\rangle + 2 c_{vv} c_{vf} \langle v^- f \rangle.
\end{eqnarray}
From squaring Eq.~\ref{eq:posit_stepC} it follows similarly that
\begin{equation}
\label{eq:posit_squared}
\langle v^- x \rangle = \frac{c_{xv}}{2} \langle v^2 \rangle.
\end{equation}
Inserting Eq.~\eqref{eq:posit_squared} into Eq.~\eqref{eq:veloc_squared} for the harmonic oscillator, $f_n = -k x_n$, and sorting coefficients then gives
\begin{equation}
\label{eq:sum_rule_ho_simple}
(1-c_{vv}^2) \langle v^2 \rangle  =  c_{vg}^2 + c_{vx}^2 \left( \frac{1-c_{vv}}{1+c_{vv}} \right) \langle x^2 \rangle.
\end{equation}
Demanding that $c_{vg}$ should not depend on the employed interaction potential, 
\begin{equation}
\label{eq:c_vg_conv_lang}
c_{vg}^2 = \left(1-c_{vv}^2\right) \frac{k_BT}{m}
\end{equation}
is the choice to be made for $c_{vg}$ as it leads to the desired kinetic energy in thermal equilibrium in the absence of an external potential. 
However, this means that the velocity variance for a harmonic oscillator will exceed its target value at any finite $\tau$, because the second summand on the r.h.s. of Eq.~\eqref{eq:sum_rule_ho_simple} is positive, since $c_{vv} < 1$ at any finite damping. 

We will finish this section by computing $\langle x^2 \rangle$.
To this end, we multiply Eq.~\eqref{eq:veloc_stepC} with $x_n$, from where
\begin{equation}
\langle v^-x \rangle = \frac{-c_{vx}}{1+c_{xv}} \langle x^2 \rangle
\end{equation}
follows.
This in turn allows the relation between $\langle x^2 \rangle$ and $\langle v^2 \rangle$  to be determined to
\begin{equation}
\label{eq:ratio_v2_x2}
\langle v^2 \rangle = \frac{-2 c_{vx}}{(1+c_{vv}) c_{xv}} \langle x^2 \rangle.
\end{equation}
Combining this relation with Eqs.~\eqref{eq:sum_rule_ho_simple} and \eqref{eq:c_vg_conv_lang} and solving for $\langle x^2\rangle$ gives
\begin{eqnarray}
\langle x^2 \rangle & = & \frac{(1+c_{vv})^2}{2+2c_{vv}-k c_{xv}c_{vf}} \frac{c_{xv}}{m c_{vf}} \frac{k_BT}{k} .
\end{eqnarray}
%
%
$\langle x^2 \rangle$ diverges when the denominator of this equation is zero, which happens when
\begin{equation}
2 ( 1+ c_{vv}) = \omega_0^2 c_{xv} c_{va},
\end{equation}
where $\omega_0^2 = k/m$, while $a$ is the acceleration from conservative forces, i.e.,  $a = f/m$ so that $c_{va} = c_{vf}/m$.
%

%

\subsubsection{Lowest-order Langevin thermostat}

In the lowest-order Langevin thermostat (LOL), the coefficients $c_{vf}$ and $c_{xv}$ are used as in the leapfrog Verlet algorithm, i.e.,
$c_{vf} = \Delta t/m$ and $c_{xv} = \Delta t$. 
The damping force is integrated as in a (non-symplectic) Euler scheme, which means that 
\begin{equation}
c_{vv} = 1 - \Delta t / \tau.
\end{equation} 
When inserting this into Eq.~\ref{eq:c_vg_conv_lang}, one can immediately recognize that $\Delta t$ must be less than $2 \tau$, as $c_{vg}$ would be imaginary otherwise. 
%

Using Eq.~\eqref{eq:ratio_v2_x2} gives the following ratio of mean kinetic and mean  potential energy
\begin{equation}
\label{eq:kin_vs_pot_lol}
\frac{U_\text{kin}}{U_\text{pot}} = \frac{2}{2-\Delta t / \tau}.
\end{equation}

To leading order
\begin{equation}
\left\langle x^2 \right\rangle = \frac{1+c_{vv}}{2} \frac{k_BT}{k}.
\end{equation}
Since, it is arguably more important to properly sample positions than velocities, it is best to re-assign $c_{vg}$ such that harmonic modes have no more leading errors linear in $\Delta t/\tau$ but instead in $\omega_0^2\Delta t^2$.
This can be achieved with
\begin{equation}
c_{vg} = \sqrt{2k_BT \Delta t/\tau},
\end{equation}
which may have been the intuitive choice to begin with, but thermal averages only happen to be so accurate due to fortuitous error cancellation and moreover increases the stability range. 
No conventional Langevin thermostat can have systematically smaller relative leading errors in the mean potential energy of harmonic oscillators than  $\omega_0^2 \Delta t^2/4$  but at best smaller higher-order errors.

\subsubsection{Best-possible conventional Langevin thermostat}

The LOL produces acceptable thermal averages at small time steps with errors in $\langle x^2 \rangle$ for the harmonic oscillator of order $\Delta t^2$. 
However, it does not lead to accurate deterministic trajectories and higher-order errors grow quickly with $\Delta t$ when the damping is large.
The question arises what the optimum choice for a conventional Langevin thermostat may be.
We would argue that the best-possible conventional Langevin (BPCL) obeys as many limiting cases or sum or product rules as possible.
Since there are four coefficients to be fixed, four independent requests can be made as long as they are not mutually exclusive.

First, the BPCL thermostat produces the exact free-particle solution for a zero external force at zero temperature, i.e.,
\begin{equation}
\label{eq:free_particle_constraint}
\frac{x_{n+1}-x_n}{x_{n} - x_{n-1}} = e^{-\Delta t/\tau},
\end{equation}
which leads to 
\begin{equation}
\label{eq:c_vv_exact}
c_{vv} = \exp(-\Delta t/\tau).
\end{equation}
Schneider and Stoll~\cite{Schneider1978PRB} already used this coefficient and it is still used in the Grønbech-Jensen thermostat~\cite{GronbechJensen2019MP}.

Second, the steady-state drift of the positions should be exact when a constant, external force $f = f_n$ is applied, i.e., the scheme should converge to
\begin{equation}
x_{n+1} = x_n + \frac{\tau \Delta t}{m} \, f
\label{eq:drift}
\end{equation}
at large $n$.
Owing to Eq.~\eqref{eq:veloc_stepC},
this is achieved when the (pseudo-) steady-state drift velocity, $v_\text{pd}$, is
\begin{equation}
v_\textrm{ps} = \frac{c_{vf}}{1-c_{vv}} f.
\end{equation}
The term \textit{pseudo} is used to indicate that the real drift velocity must be determined through the time evolution of $x(t)$ rather than by time-averaging $v(t)$.
Using this result for the velocity 
in Eq.~\eqref{eq:posit_stepC} and equating predicted and exact drifts provides the constraint
\begin{equation}
\label{eq:xv_vf_constraint}
c_{xv} c_{vf} = (1-c_{vv}) \frac{\tau \Delta t}{m}.
\end{equation}

Third, mean kinetic and potential energy assume the same value in true thermal equilibrium.
Exploiting Eq.~\eqref{eq:posit_squared} and solving for $c_{xv}$ nd $c_{vf}$ yields
\begin{eqnarray}
c_{xv} & = & \sqrt{2\tau\Delta t \frac{1-c_{vv}}{1+c_{vv}}} \\
c_{vf} & = & \sqrt{\tau\Delta t (1-c_{vv}^2)/2} / m.
\end{eqnarray}

Fourth, the mean kinetic energy of a thermostatted but otherwise force-free point mass should be exact.
This leads again to Eq.~\eqref{eq:c_vg_conv_lang} so that
\begin{equation}
\langle x^2 \rangle = \frac{2+2c_{vv}}{2 + 2c_{vv} - \omega_0^2 (1-c_{vv})\tau\Delta t} \frac{k_B T}{m}.
\end{equation}
Thus, this time, $c_{vg}^2$ does not require a ``fudge factor'' to have the leading relative errors in the mean potential energy of the harmonic oscillator be $\omega_0^2\Delta t^2/4$.
In fact, it might have been more meaningful to demand that $c_{vg}$ is chosen such that $\langle x^2\rangle$ has the smallest possible model-independent, i.e., $\omega_0$-independent error, as the positions are ``true'' and not ``pseudo''.

The issue of pseudo velocities reappears in the GJ thermostat but can be discussed already here in the context of the BPCL. 
Many observables of interest depend on velocity. 
Since it is not generally possible to discriminate between a pseudo-drift velocity and thermal fluctuations, it is best to deduce velocity-dependent observables from the actual trajectories $x(t=n\Delta t) = x_n$ through numerical differentiation, Fourier analysis, or a related technique.
Of frequent interest is the mass-weighted (MW) velocity autocorrelation (ACF)
\begin{equation}
C_{vv}^\text{MW}(\tau)  = \sum_n m_n \langle \mathbf{v}_n(t) \cdot \mathbf{v}_n(t+\tau) \rangle
\end{equation}
as it allows spectral properties to be deduced but also quantum corrections to the specific heat to be estimated~\cite{Gao2021PRM}. 
Observables like $C_{vv}^\text{MW}(\Delta t)$ can be computed indirectly from corresponding correlation functions of the positions, in this case from the total mass-weighted mean-square-displacement (MSD)
\begin{equation}
\text{MSD}_\text{MW}(\tau)  =  
\sum_n m_n \left\langle 
\left( \mathbf{r}_n(t+\tau)  - \mathbf{r}_n(t) \right)^2
\right\rangle
\end{equation}
via
\begin{equation}
C_{vv}(\tau) = \frac{1}{2} \frac{d^2}{d\tau^2} \text{MSD}_\text{MW}(\tau).
\end{equation}
Hence, in the following, the focus will be on position-dependent observables.
Any velocity-dependent observable can be deduced indirectly.

\subsection{Advanced Stochastic Thermostats}

Exact thermal averages for harmonic references can be obtained when using two random numbers in one time step, where one is a new random number and the other is recycled from the previous time step. 
Corresponding thermostats will be motivated again by defining the integration scheme and making them obey certain asymptotic limits.
%
One thermostat implements Brownian dynamics, which is produced in the limiting case of the Langevin equation, in which $m$ is send to zero, while damping, $\gamma = m/\tau$, is kept constant. 
The other is the solver of the Langevin equation proposed by Grønbech-Jensen (GJ)~\cite{GronbechJensen2019MP}.

\subsubsection{Advanced Brownian thermostat}
\label{sec:brownian}


A highly accurate solver for Brownian dynamics, proposed by Grønbech-Jensen and Farago (GJF) can be constructed from the integration scheme~\cite{GronbechJensen2013MP}
\begin{eqnarray}
x^{++} & = & x + c_{xf} f + c_{xg}(g^{+} + g^{-}),
\label{eq:brown_osc}
\end{eqnarray}
where we use similar notation as above, e.g., $x^{++}$ stands for $x_{n+1}$, $f$ for $f_n$ and $g^-$ for 
$g_{n-1/2}$. 
For a harmonic oscillator, the integration scheme reduces to 
\begin{eqnarray}
x^{++}  & = & (1-c_{xf}k) x + c_{xg}(g^+ + g^-).
\label{eq:brown_osc_ho}
\end{eqnarray}
Obtaining the correct asymptotic drift under a constant force requires $c_{xf}$ to be
\begin{equation}
c_{xf} = {\Delta t}/{\gamma}
\end{equation}
%
so that
\begin{equation}
\langle x g^- \rangle = c_{xg}.
\label{eq:xg_corr_brown}
\end{equation}

Squaring both sides of Eq.~\eqref{eq:brown_osc_ho} and taking the expectation value while using Eq.~\eqref{eq:xg_corr_brown} yields
\begin{equation}
\langle x^2\rangle = (1 - c_{xf} k )^2 \langle x^2\rangle
+ 2 (1 - c_{xf} k ) c_{xg}^2 
+ 2 c_{xg}^2,
\end{equation}
which can be solved to give
\begin{equation}
c_{xg}^2 = { k_B T \Delta t}/ (2{\gamma})
\end{equation}
when requesting that $\langle x^2 \rangle = k_BT/k$.

\subsubsection{Grønbech-Jensen thermostat}
\label{sec:gronbech}

In our notation, the GJ algorithm~\cite{GronbechJensen2019MP,GronbechJensen2013MP} can be written as
\begin{subequations}
\label{eq:gj_scheme}
\begin{eqnarray}
\label{eq:veloc_step}
v^{+} & = & c_{vv} v^{-} + c_{vf} f + c_{vg}(g^{-}+g^{+}) \;\;\;\; \\
\label{eq:posit_step}
x^{++} & = & x + c_{xv} v^{+},
\end{eqnarray}
\end{subequations}
%
%
As in the advanced Brownian integrator, random numbers are added to the velocity in two successive rather than in a single  step. 
Thus, only those previous results can be recycled that do not depend on thermal random forces, most notably Eqs.~\eqref{eq:posit_squared}, \eqref{eq:c_vv_exact}, and \eqref{eq:xv_vf_constraint}.
Other relations must be rederived, in particular the result for $\langle v^2 \rangle$ obtained from squaring the velocity time stepping equation, Eq.~\eqref{eq:veloc_step}:
\begin{eqnarray}
\label{eq:general_covariance}
(1-c_{vv}^2) \langle v^2 \rangle & = & c^2_{vf} k^2 \langle x^2 \rangle + 2 c_{vg}^2 -
2c_{vv}c_{vf}k \langle v^- x\rangle \nonumber\\
& & + 2c_{vv} c_{vg} \langle v g \rangle - 2 c_{vf} c_{vg} k \langle x g^-\rangle.
\end{eqnarray}

To proceed, we need the covariances $\langle v g\rangle$ and $\langle x g^-\rangle$.
They can be obtained 
multiplying Eq.~\eqref{eq:veloc_step} with $g^+$ and by taking expectation values on both sides 
\begin{equation}
\langle v g \rangle = c_{vg}.
\end{equation}
This in turn leads to
\begin{equation}
\langle x g^- \rangle = c_{xv} c_{vg}.
\end{equation}
Ideally, the integration scheme 
satisfies equipartition
\begin{subequations}
\label{eq:equi_partition}
\begin{eqnarray}
\langle x^2 \rangle & = & k_BT/k \\
\langle v^2 \rangle & = & k_BT/m .
\end{eqnarray}
\end{subequations}
Inserting these results into Eq.~\eqref{eq:general_covariance} and rearranging terms leads to
\begin{eqnarray}
\label{eq:condition_gj}
0  &  = & 
 k \left( c_{vf}^2 k_BT - c_{vv} c_{vf} c_{xv} {k_BT}/{m} - 2 c_{vf} c_{xv} c_{vg}^2 \right)
\nonumber\\
& &  - (1-c_{vv}^2) {k_BT}/{m} + 2 (1+c_{vv}) c_{vg}^2  .
\end{eqnarray}
Eq.~\eqref{eq:condition_gj} should hold irrespective of the value of $k$.
Thus,
\begin{equation}
\label{eq:c_vg_gj}
c_{vg} = \sqrt{(1-c_{vv}) k_BT / (2m)}
\end{equation}
and later
\begin{equation}
c_{vf} = c_{xv} / m. 
\end{equation}
Combining this last equation with Eq.~\eqref{eq:xv_vf_constraint} then gives 
\begin{eqnarray}
c_{xv} & = & \sqrt{(1-c_{vv})\tau\Delta t} \nonumber\\
c_{vf} & = & c_{xv}/m.
\end{eqnarray}

Up to this point, we have only shown that the $c_{\alpha\beta}$ are optimal provided that 
Eq.~\eqref{eq:equi_partition} is a self-fulfilling prophecy. 
That this is indeed the case can be confirmed by evaluating all possible covariances that can be deduced using Eqs.~\eqref{eq:veloc_step} and \eqref{eq:posit_step} and ensuring that the resulting equations do not conflict with each other. 
A more practical approach is to implement the algorithm and verify that the second and potentially higher moments converge to their expected values as the statistics improve.

As a final remark on the GJ thermostat, we note that instantaneous thermal fluctuations turn out to be \emph{true fluctuations} in the sense that their distribution is exact for a harmonic reference~\cite{Farago2019PA}, in our case by construction.
We would nevertheless call the velocities \emph{pseudo}, because the steady-state drift velocity obtained by averaging the instantaneous velocity using the GJ integrator differs from the exact value.

\subsection{Rheological models and colored thermostats}

In many cases, there is the desire to solve a modified Langevin equation, where damping is not instantaneous but occurs with a delay. 
One reason can be that the computation of the inter-atomic forces require an on-the-fly computation of auxiliary degrees of freedoms (DOFs), such as wavefunction coefficients in Car-Parrinello MD~\cite{Car1985PRL} or atomic dipoles or charges in classical potentials~\cite{Muser2022APX}. 
Extended Lagrangian schemes have proven useful in this endeavor, the idea being that the auxiliary DOFs fluctuate around their smoothly evolving target values while approaching it. 
This can obviate the need for a brute-force minimization of energy, enthalpy, or related expressions with respect to the auxiliary DOFs.  
However, the erratic, short-time motion of the coordinates induced by short-time thermal noise is obviously disadvantageous.
Delaying the random signal through a memory-functional formalism
\begin{equation}
\label{eq:general_langevin}
m\ddot{x} + \int_{-\infty}^t \textrm{d}t' \, \eta(t-t') \, \dot{x} = f(x) + \Gamma(t),
\end{equation}
where the ACF of the random noise satisfies the fluctuation-dissipation theorem
\begin{equation}
\langle \Gamma(t) \Gamma(t') \rangle = 2k_BT \eta(t-t')/m
\end{equation}
substantially reduces short-time erraticity~\cite{Ceriotti2009PRL,Ceriotti2010JCTC}. 

\begin{figure}[thb]
\begin{center} 
\includegraphics[width=0.45\textwidth,angle=0]{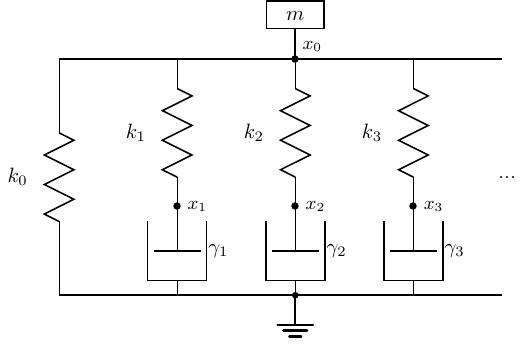}
\caption{\label{fig:maxwell}
Coupling of a central harmonic degree of freedom to a series of Maxwell models. 
The coupling of the mass to the spring of stiffness $k_0$ represents the deterministic force. 
All other couplings form the thermostat.
} 
\end{center}
\end{figure}

Another reason for solving equations like Eq.~\eqref{eq:general_langevin} is to actually mimick true dynamics, for instance, when modeling viscoelastic response functions.
This is often done by considering the athermal limit of Eq.~\eqref{eq:general_langevin} and by representing the memory kernel through Maxwell models~\cite{Bugnicourt2017TI,Sukhomlinov2021ASSA}, as depicted in Fig.~\ref{fig:maxwell}.
When exposing the Maxwell element $x_n(t)$, $n\ge 1$ to Brownian or Langevin dynamics, they turn into explicitly treated bath variables driving the central DOF to their temperature, whereby they produce similar dynamics as if their effect was considered implicitly as in the colored thermostat proposed by Ceriotti, Bussi, and Parrinello~\cite{Ceriotti2009PRL,Ceriotti2010JCTC}.

%
Depending on the algorithm, direct simulation of Maxwell elements can suffer from typical thermostat issues: inaccurate $\langle x^2 \rangle$ and small time steps for strong damping.

Here, we propose to subject the thermostat variables, $x_{n>1}(t)$, to Brownian dynamics as described in Sect.~\ref{sec:brownian}, instead of directly thermostatting the central DOF.
As the GJF thermostat is applied to a Maxwell element, the method will be denoted by GJFM. 
%
The extra degree of freedom does not cause memory overhead compared to using a low-pass filtering technique, since the low-pass filtered random force no longer needs to be retained. 
At the same time, the propagator will be exact for harmonic references and hence can be expected to perform with a similar precision for close-to-harmonic systems as the bare GJ. 

The typical parameterization of Maxwell (Mxw) elements’ damping and spring coefficients varies depending on whether they are used to (i) thermostat a DOF coupled to energy-minimizing auxiliary parameters, (ii) directly thermostat auxiliary DOFs, or (iii) model thermal fluctuations in coarse-grained viscoelastic systems.
In the first case, the Maxwell element’s spring constant should generally not exceed the natural (or environment-dependent) stiffness of the thermostatted DOF to prevent high-frequency thermal forces. 
By contrast, when modeling rheological response functions, larger spring constants are used, since elastomers exhibit high-frequency elastic moduli that far exceed their quasi-static values.

\begin{figure}[hbtp]
    \centering
    \includegraphics[width=0.95\linewidth]{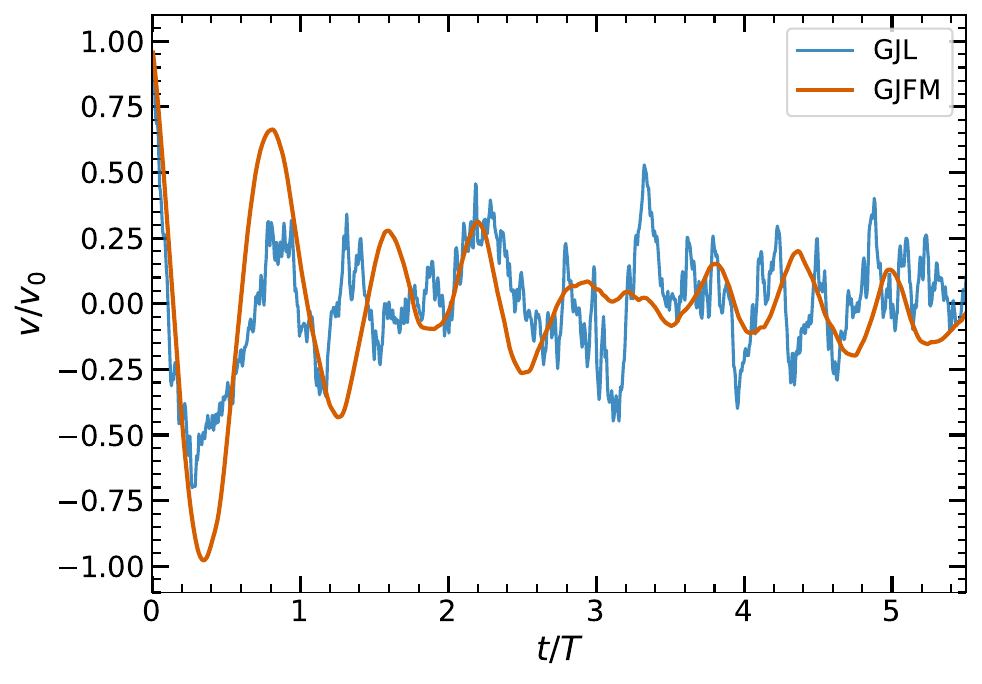}
    \caption{
    Velocity $v$ as a function of time $t$ for a harmonic oscillator with stiffness $k=1$ and mass $m=1$ at thermal energy $k_B T = {m v_0^2}/{20}$, where $v_0 = 1$ is the initial velocity.  
    Data for the GJL thermostat is plotted as a thin blue line, while data for the GJFM thermostat is shown as a thick orange line.  
    The GJFM uses a single Maxwell element.  
    Both thermostats have damping $\gamma = 1$.  
    The spring stiffness for the GJFM is set to $k_\text{Mxw} = k/2$. 
}
    \label{fig:vel_vs_t_mxw}
\end{figure}

To compare the GJFM method with the other thermostats examined, the Maxwell damping is chosen to reproduce the steady-state velocity damping, i.e., $\gamma_\text{Mxw} = m/\tau$. 
The Maxwell (``stress'') relaxation time, $\tau_\text{Mxw} \equiv k_\text{Mxw}/\gamma_\text{Mxw}$, is set to an estimate of the Debye time, $T_0 = 2\pi/\omega_0$, which represents a characteristic (or minimal) period of the system within the harmonic approximation. 
Of course, the formal solution and general properties of the Maxwell thermostat do not depend on the specific parameterization. 
Fig.~\ref{fig:vel_vs_t_mxw} compares normalized velocity relaxation profiles for a single harmonic degree of freedom subject to an initial velocity $v_0$, thermostatted using the GJL and single-element GJFM schemes.
Despite having comparable long-time damping characteristics, the two schemes differ in how they transmit thermal noise at short times.
In contrast to the noisier velocity fluctuations in GJL, the GJFM exhibits smoother and more structured damping, consistent with the memory effects introduced by the Maxwell element.

Although, the formal solution of the equations of motion resulting from Fig.~\ref{fig:maxwell} should  be textbook material, we are not aware of such a text.
This is why we quickly present it here with the added benefit of showing the validity of the fluctuation-dissipation theorem~\cite{Kubo1966RPP} for the proposed thermostatting scheme.
The full set of Langevin equations for the system depicted in
Fig.~\ref{fig:maxwell} is
\begin{subequations}
\label{eq:fullSetMaxwell} 
\begin{eqnarray}
\label{eq:fullSetMaxwella}
m \ddot{x}_0 + k x_0 & = & \sum_n k_n\, (x_n - x_0) + F_{\rm ext}(t) \\
\eta_n \dot{x}_n & = & k_n\, (x_0 - x_n) + \Gamma_n(t) ,
\end{eqnarray}
\end{subequations}
where $F_{\rm ext}(t)$ is an external force, and where the $\Gamma_n(t)$
obey Eq.~\eqref{eq:random_force} for each $x_n$ individually. 

Let us first define an effective, frequency-dependent damping, which
is obtained by ignoring the random noise, but by keeping the external
force.
To do that, the equations of motion are expressed in terms of their
Fourier transforms so that the $\tilde{x}_n$ can be eliminated:
\begin{equation}
\left\{ -m \omega^2 - i \omega \eta(\omega) + k_0\right\}
\tilde{x}(\omega) = \tilde{F}_{\rm ext}(\omega)
\label{eq:xOfFinFourierMaxw1}
\end{equation}
where
\begin{equation}
\eta(\omega) = \sum_n \frac{\eta_n}{1-i\tau_n \,\omega}
\end{equation}
and $\tau_n = \eta_n/k_n$.
$\eta(\omega)$ is nothing but the Fourier transform of 
\begin{equation}
\eta(t) = \Theta(t) \, \sum_n \eta_n e^{-t/\tau_n}
\end{equation}
where $\Theta(t)$ is the Heaviside step function.
Next, the set Eq.~(\ref{eq:fullSetMaxwell}) is used to compute the
ACF of the forces due to the Maxwell models, this time, however,
in the presence of thermal noise but without external force, whose effect has already been accounted for. 
The formal solutions for the $\tilde{x}_n(\omega)$ can be
obtained as a function of the $\tilde{\Gamma}(\omega)$ and $\tilde{x}_0$.
Inserting them into Eq.~(\ref{eq:fullSetMaxwella}) yields an
equation in which $\tilde{F}_{\rm ext}(\omega)$ on the r.h.s. of
Eq.~(\ref{eq:xOfFinFourierMaxw1}) must be replaced with
\begin{equation}
\tilde{\Gamma}(\omega) = 
\sum_n \frac{1}{1-i\omega \tau_n} 
\tilde{\Gamma}_n(\omega)
\end{equation}
so that
\begin{equation}
\left\langle
\tilde{\Gamma}^*(\omega)
\tilde{\Gamma}(\omega')
\right\rangle =
\sum_n 
\frac{2k_BT \eta_n \delta(\omega-\omega')}{1+(\omega \tau_n)^2},
\end{equation}
since the ``usual rules'' apply for the $\tilde{\Gamma}_n$ with $n > 0$.
Using the Wiener-Khintchine theorem, the ACF in the real-time domain
reads
\begin{equation}
\langle \Gamma(t) \Gamma(t') \rangle = 
2 k_B T \sum_n \eta_n \exp(-t/\tau_n),
\end{equation}
which, except for a prefactor,  is nothing but the time-dependent damping.

\subsection{Momentum-conserving Langevin thermostat}

Dissipative-particle dynamics (DPD) is supposedly the first stochastic momentum-conserving thermostat~\cite{Espanol1995EPL,Vattulainen2002JCP}.
In DPD, two nearby particles interact hydrodynamically via central damping and random forces in their center-of-mass system, whereby the effect of unaccounted DOFs is incorporated approximately.  
%
By construction, neither damping nor random forces act on the center of mass of the two particles 
so that DPD is not only globally but also locally momentum conserving. 

While DPD certainly is extremely use- and powerful and still relatively straightforward to code, it is more difficult to implement than Langevin thermostats acting in a frame of reference. 
Moreover, it requires a large number of random numbers to be drawn, that is, one for each pair of hydrodynamically interacting particles.
Last but not least, it requires small time steps to be taken, when the local damping is large. 

Extending the GJ or even the BPCL thermostat to DPD bears the potential to use large local damping, while maintaining a large time step.
However, the free-particle propagator no longer factorizes into the product of single-particle propagators, so that a naive porting of the algorithms, in which the propagation coefficients keep being scalars, will not obey constraints like those obtained in Eqs.~\eqref{eq:free_particle_constraint} and \eqref{eq:drift}.
Moreover, the damping coefficients generally depend on the interparticle distances and thus changes with time.
This situation is not accounted for in the derivation of the accurate Langevin schemes. 

It \textit{might} be possible to overcome some problems with a diagonalization procedure, which would have to be conducted at each time step.
However, we do not see an easy solution for systems with time dependent damping, which automatically arise, when the topology of atoms interacting with DPD thermostats can change with time. 
Exceptions arise when the eigenmodes can be computed efficiently via fast (Fourier) transforms, for example, crystals where damping coefficients are defined by topology rather than by distance.  
A particularly appealing, though somewhat niche, application of this approach would be boundary-element methods for describing the contact mechanics of elastomers.  
In that case, damping coefficients can depend on the wavelength of surface undulations to mimic the rheology of the underlying linear viscoelastic body~\cite{Sukhomlinov2021ASSA}, with a Fourier-based solution best propagated using GJF.

As an alternative to DPD, we propose to subdivide the simulation cell into sub-volumes or (binning) boxes.
An appropriate linear dimension should be of order two atomic diameters so that two atoms find themselves in a linear system and a few more, say eight atoms total, in three spatial dimensions, $D = 3$.
In each box, the motion of atoms relative to the box's center of mass is subjected to an (advanced) thermostat, while the center of mass of all atoms in the box are propagated without thermostatting using a symplectic integration scheme.  
A more detailed pseudo-code of the scheme, in which is given next.
We hope that the nomenclature is clear and sufficiently self-explanatory, e.g., \texttt{x}, \texttt{v}, \texttt{f}, and \texttt{g} retain their meanings from previous equations or listings and all variables in the inner loop, including the center-of-mass variables (indicated by \texttt{\_com}) and other means, relate to the given box. 
The actual implementation is provided in the supplementary material. 

\begin{algorithmic}
\STATE \textbf{def} \texttt{propagator(self, x, v, f):}
\STATE \quad draw random origin and define binning boxes
\STATE \quad bin atoms into boxes
\STATE \quad \textbf{for each box:}
\STATE \quad\quad compute \texttt{x\_com}, \texttt{v\_com}, \texttt{f\_com}
\STATE \quad\quad subtract \texttt{x\_com}, \texttt{v\_com}, \texttt{f\_com} from \texttt{x}, \texttt{v}, \texttt{f} 
\STATE \quad\quad \texttt{g = g - mean(g)} for \texttt{g} in bin
\STATE \quad\quad propagate particles using BPCL
\STATE \quad\quad propagate \texttt{x\_com}, \texttt{v\_com} with a sympletic scheme
\STATE \quad\quad add \texttt{x\_com}, \texttt{v\_com}, \texttt{f\_com} to \texttt{x}, \texttt{v}, \texttt{f}
\end{algorithmic}

We are aware that the approach can be streamlined, e.g., subtracting and adding \texttt{x\_com} is not really needed.
Moreover, in each bin, only $D\times (N_b-1)$ random numbers need to be drawn, insted of $D N_b$, where $D$ is the spatial dimension and $N_b$ the number of atoms in a bin $b$.
However, these few unnecessary operations were kept for the sake of readability.

The approach is obviously momentum conserving but, unlike DPD, not angular-momentum conserving so that viscosity will be more enhanced in this scheme than in DPD, given similar damping constants. 
Yet, a single particle in a box would remain unthermostatted; ideal gas atoms would be ballistic until they happen to find themselves in a bin with at least one other atom. 

In principle, all Langevin thermostats can be easily applied to the proposed momentum conserving stochastic thermostat. 
However, the accuracy of advanced thermostats can be decremented, since obeying asymptotic limits like those of Eqs.~\eqref{eq:free_particle_constraint} and \eqref{eq:drift} require again the global free-particle propagator to be diagonalized. 
Despite being a tedious procedure, it would nevertheless be a substantially simpler exercise now than for DPD, since the thermostat no longer act parallel to time-dependent directions but simply parallel to all spatial directions. 

The issue of violating asymptotic limits appears to be more problematic for GJ than for BPCL when randomly choosing the origin of the coordinate system for the placement of the binning boxes. 
In fact, in our analysis of a linear LJ chain with random binning box assignment, BPCL clearly outperformed GJ.
Since particles would move in and out of binning boxes in the overwhelming majority of cases, we decided not to explore the GJ-based momentum conserving thermostats in this study any further.
%
%

\subsection{Global-kinetic-energy controls}
\label{sec:tot_t_kin_ctrl}

In this paper, occasional comparison will be made to schemes in which the velocities of all particles in the cell is rescaled by the same factor such that the first moment of the kinetic energy~\cite{Berendsen1984JCP} and potentially the higher moments adopt their exact desired value after equilibration.
The goal is to minimally perturb Newtonian dynamics while still ensuring accurate thermal averages of quantities beyond kinetic energy moments such as configurational energy or pair distribution functions.
Here, we do not intend to repeat the details of such global kinetic energy controls (GKECs), for two main reasons.
First, they have been discussed extensively in the literature, including in standard textbooks~\cite{AllenTildesley2017book,Frenkel2001book}.
Second, and more importantly, as we argue further below, velocity rescaling and thermostatting are mutually exclusive in simple harmonic reference systems. This may appear artificial at first, but such systems are, upon closer inspection, nearly ubiquitous.
This is a well-known feature of purely harmonic systems~\cite{Patra2014PRE}.
However, it may be useful to highlight this issue in the context of quasi-harmonic modes, which inevitably arise in sufficiently large, homogeneous systems, and to provide general arguments explaining why such methods fail in these cases.

GKECs fail when two identical and similarly initialized sub-systems, or, certain (collective) DOFs, are part of the total system and couple to the remaining DOFs at best marginally, then their motion will be fully correlated for the remainder of the simulation.
This means any GKEC will not recognize that the real coupling of the two sub-systems is weak.
As such, GKECs can never guarantee to produce a canonical distribution, i.e., they cannot be thermostats in the proper sense.

To make our points formal, assume a many-particle system, in which two degrees of freedom (DOFs), $x_1(t)$ and $x_2(t)$  happen to be harmonic but essentialy uncoupled to the rest of the system.
If these two DOFs happen to have the same inertia $m$ and the same eigenfrequency $\omega_0$ then they will remain correlated for the rest of eternity when subjected to a global and/or deterministic velocity rescaling (vr) scheme.
Thus, the predicted correlation matrix satisfies
\begin{equation}
\langle x_i(t) x_j(t) \rangle_\textrm{vr} =
\frac{k_B T }{m \omega_0^2}
\cos\varphi_{ij},
\end{equation}
where $\varphi_{ij}$ is the initial phase shift betweem the oscillators $i$ and $j$, which is not going to change when rescaling is global and/or deterministic.
This differs from the exact (ex) relation
\begin{equation}
\langle x_i(t) x_j(t) \rangle_\textrm{ex} =
\frac{k_B T }{m \omega_0^2} \delta_{ij}.
\end{equation}
Even if the two DOFs are assigned different initial velocities, their phases will never cross — which also violates the canonical distribution.

The example is far from being constructed.
Any large simulation cell containing a homogeneous system will have pairs of long wavelength modes that are degenerate, i.e., density oscillations of the form $\cos(2\pi x/L_x)$ and $\sin(2\pi x/L_x)$, where $L_x$ is the length of the simulation box in $x$-direction.
Since any homogeneous system is asymptotically harmonic, the two modes will take the longer to decohere the larger the system.
Another situation is a simulation in a constant stress ensemble, where the box shape variables are the propagated DOFs.
Analysis of the box fluctuations allows the elastic tensor to be reconstructed.
In an isotropic system with an initial cubic symmetry, the diagonal strain tensor elements will, in most cases have, the same initial condition.
Since non-elastic coupling of internal degrees of freedom to the box shape are very weak, in particular for large system, the determination of the compliance tensor
\begin{equation}
S_{\alpha\beta\gamma\delta} = \frac{V}{k_BT} \langle \varepsilon_{\alpha\beta} \varepsilon_{\gamma\delta} \rangle
\end{equation}
turns out flawed.
This may be one reason why it appears difficult to identify works reporting the full elastic tensor from simulations using velocity-rescaling algorithms for barostatting, while Langevin-based thermostats can yield quite reasonable estimates already after a few hundred time steps, provided that good values for the box inertia and damping were employed.
In fact, it will probably be best to thermostat box shapes with GJF and to use different damping for diagonal and off-diagonal elements of the $h$-matrix, e.g., proportional to the bulk and shear moduli, respectively, to quickly obtain independent observations.
Our analysis above also explains why noise-cancellation techniques for computing elastic constants at finite temperature tend to yield more accurate results when using Langevin-based thermostats rather than GKEC-based methods~\cite{Mukherji2025JCP}.

\subsection{The linear harmonic chain}
\label{sec:harmonic_chain}
The equation of motion of the harmonic chain in the absence of an external force reads
\begin{align}
\label{eq:linear_chain_atomic}
& m \ddot{u}_i + \frac{m}{\tau_\text{l}}\dot{u}_i + \left(k + \frac{\mu}{\tau_\text{r}}\frac{\partial}{\partial t}\right) (2{u}_i - {u}_{i+1} - u_{i-1}) =\nonumber  &\\ 
&  \Gamma_{\text{l}i}(t) + \Gamma_{\text{r}i,i\pm 1}(t) .&
\end{align}
Here, $u_i$ is the displacement of the atom or discretization point $i$, $m$ is its mass, while $\mu$ is the reduced mass of two neighbors.
The index l stands for the laboratory frame of reference and r for relative motion, while the $\tau_{r,l}$ are damping times. 
The random force acting in the laboratory system, $\Gamma_{\text{l}i}(t)$, is the same as that introduced in Sect.~\ref{sec:langevin}. 
The ``relative'', momentum-conserving random forces, $\Gamma_{\text{r}i,i+1}(t) = -\Gamma_{\text{r}i+1,i}(t)$, also have a first moment of zero.
Their autocorrelation function is given by
\begin{eqnarray}
\left\langle \Gamma_{\text{r}i,j}(t)  \Gamma_{\text{r}i,j}(t') \right\rangle& = 2 \delta_{i,j\pm 1} \mu k_B T  \delta(t-t') / \tau_\text{r}.
\end{eqnarray}
Given an equilibrium distance $a$ between two adjacent atoms and substituting, e.g., $u_i \to u(ia)$ and $u(x+a) - u(x) \to au'(x+a/2)$, Eq.~\eqref{eq:linear_chain_atomic} can be cast in the continuum approximation at $k_BT = 0$ as
\begin{equation}
{\ddot{u}} + \frac{1}{\tau_\text{l}} {\dot{u}} 
+  \left(\omega_0^2 +  \frac{\mu}{m \tau_\text{r}}  \frac{\partial}{\partial t}  \right) a^2 \Delta u  = 0
\end{equation}
with $\omega_0^2 ={k}/{m}  $
The solution to this homogeneous partial differential equation can be written as the superposition of waves $\tilde{u}(q,\omega) \exp\left\{ \text{i}(qx-\omega t)\right\}$, each of which must satisfy
($\partial/\partial x \to \text{i}q$ and $\partial/\partial t \to -\text{i}\omega$)
\begin{equation}
\omega^2  + \text{i}\omega / {\tau_\text{l}} +
q^2 a^2 \left( \omega_0^2 - \text{i} \omega / \tau_\text{r}\right) = 0.
\end{equation}
Discreetness effects can be reintroduced by substituting $q^2a^2$ in this last equation with $4\sin^2(qa/2)$. 

Solving for $\omega$ yields
\begin{equation}
\omega(q) = -\textrm{i}\gamma(q)/2 \pm \sqrt{\omega_0(q) - \gamma^2(q)/4}
\end{equation}
with
\begin{subequations}
\begin{align}
\omega_0(q) & = \pm aq \omega_0 \\
\tau^{-1}(q) &= \tau^{-1}_\text{l} + (aq)^2 \tau^{-1}_\text{r}.
\end{align}
\end{subequations}
This results implies that local damping makes small-$q$ or large wavelengths displacements be overdamped relative to those at large-$q$, i.e., the quality factor $Q(q) = 1/(\omega(q)\tau(q))$ scales with $1/q$ for the damping in the laboratory system but proportionally to $q$ for the momentum-conserving damping.  

Thus, the relative damping and thereby the (longitudinal but also the regular, transverse) viscosity diverges in the thermodynamic limit when thermostatting relative to a frame of reference.
In contrast, momentum-conserving thermostatting adds a constant offset to the damping and thus viscosity.
In other words, hydrodynamic interactions are eliminated in the first case and reduced (in a controllable fashion) in the second. 

A consequence outside physics of this scaling is that lower frequency sound can travel further than high-pitched noise. 
This might have had practical consequences for the evolution of species communicating through sound, which do not appear to have been discussed in the pertinent literature~\cite{Puts2010EHB}. 
If one of the two sexes is particularly valuable, e.g., because they can reproduce only once per year and are thus needed after a sudden decline in population, it would be wasteful to have representatives of that sex defend territory, conquer new one, or expose them to life-threatening hunting. 
In all these latter activities it is advantageous to communicate over long distances, i.e., to have low voices.
The advantage of high-pitched voices is that more information can be conveyed per time unit over short distances.

\subsection{Used Codes}

The results presented in this work were produced using three different codes.
LAMMPS~\cite{Thompson2022CPC} was employed whenever it allowed for the desired simulations to be performed.
In addition, we developed two in-house codes, which are available for download from a publicly accessible GitHub repository.

For completeness, we note that these in-house codes were primarily developed to explore possible directions for future developments, rather than serving as the main computational tools for this study.
Specifically, one of our goals is to design a boundary-element code for the simulation of mechanical boundary conditions that is more easily extensible than current implementations.
Former implementations of Green's function molecular dynamics~\cite{Kong2009CPC} are no longer available in LAMMPS.
Furthermore, we are interested in developing charge-transfer potentials, in which the conductance of a bond in the split-charge formalism can dynamically transition between finite and zero values~\cite{Muser2024JCP}.
Thermostatting the charge-transfer variables requires an algorithm capable of seamlessly transitioning from Langevin to Brownian dynamics while still permitting large time steps.

In one case, MH developed a framework for running simulations in Rust.
This in-house code will be referred to as IHRC.
It was used to explore three-dimensional Lennard-Jonesium systems, for which the BPCL implementation was not available in LAMMPS.

The second in-house code was written in Python by MHM and will be referred to as IHPC.
SA contributed functionality for the Maxwell-element thermostat and co-developed the momentum-conserving thermostat.
A particular feature of IHPC is the ability to run different replicas in parallel, allowing for on-the-fly error estimation.
Moreover, running multiple replicas compensates for Python’s slower performance through the use of \texttt{numpy} and just-in-time compilation techniques.

SS identified and corrected various errors in the GKEC and NHC functions originally implemented in IHPC.
However, for optimal performance, SS adapted LAMMPS input files to enable simulations of the one-dimensional model systems directly within LAMMPS.

\section{Results}
\label{sec:results}

In this section, we scrutinize the thermostatting schemes described above on specific systems.  
To this end, we simulate generic, computationally lean models that allow accurate statistics to be readily obtained.  
Despite their simplicity, some of these models mimic the complexity of real systems, exhibiting phenomena such as shear thinning or dynamics resembling the $\alpha$- and $\beta$-relaxation of glass-forming melts.  
Nonetheless, the critically damped harmonic oscillator plays the most central role.  
This is because critical damping allows a \hbox{(quasi-)} harmonic mode to relax to equilibrium in a fast and efficient manner.  
This proves useful not only when simulating auxiliary DOFs but also for what we term semi-auxiliary DOFs, such as the size or shape of a simulation cell~\cite{Parrinello1981JAP}.  
The latter is a near-harmonic variable with weak coupling to the thermal fluctuations of atomic positions within the cell.  
Other examples include induced dipoles or transfer charges used in charge-equilibration approaches~\cite{Muser2022APX}.  
To achieve the fastest possible convergence for their first and second moments, it is clearly desirable to choose their inertia as small as possible so that their correlation time is minimized.  
This, in turn, reduces stochastic errors in thermodynamic properties, including linear response functions, which can be inferred from the analysis of second moments.

\subsection{Deterministic trajectories}
\label{sec:deterministic}

When initializing a thermostated system with a given initial condition deviating from thermal equilibrium, the first velocity, $v_{-1/2}$, assigned in the GJ scheme is best set to  
\begin{equation}
\label{eq:assign_v}
v_{-1/2} = \frac{\Delta t}{c_{xv}} v_\text{determ}(-\Delta t/2) + \sqrt{\frac{k_B T}{m}} g,
\end{equation}  
where $g$ is, as usual, a standard Gaussian random number, while $v_\text{determ}$ is the deterministic or expected velocity at the moment of initialization, i.e., at the time $t = -\Delta t/2$.  
This way, the position drifts in the desired deterministic manner, while deviations in the velocity reflect true thermal fluctuations.  
Of course, initial conditions are usually specified as $x(0) = x_0$ and $v(0) = v_0$.  
To obtain $v(-\Delta t/2)$, a half reverse time step must be executed, or, as we do here, the exact velocity at $t = -\Delta t/2$
is assigned for the initial condition considered in this section:  
$x(0) = 1$ and $v(0) = 0$.

\begin{figure}[h!]
\begin{center}
\includegraphics[width=0.45\textwidth]{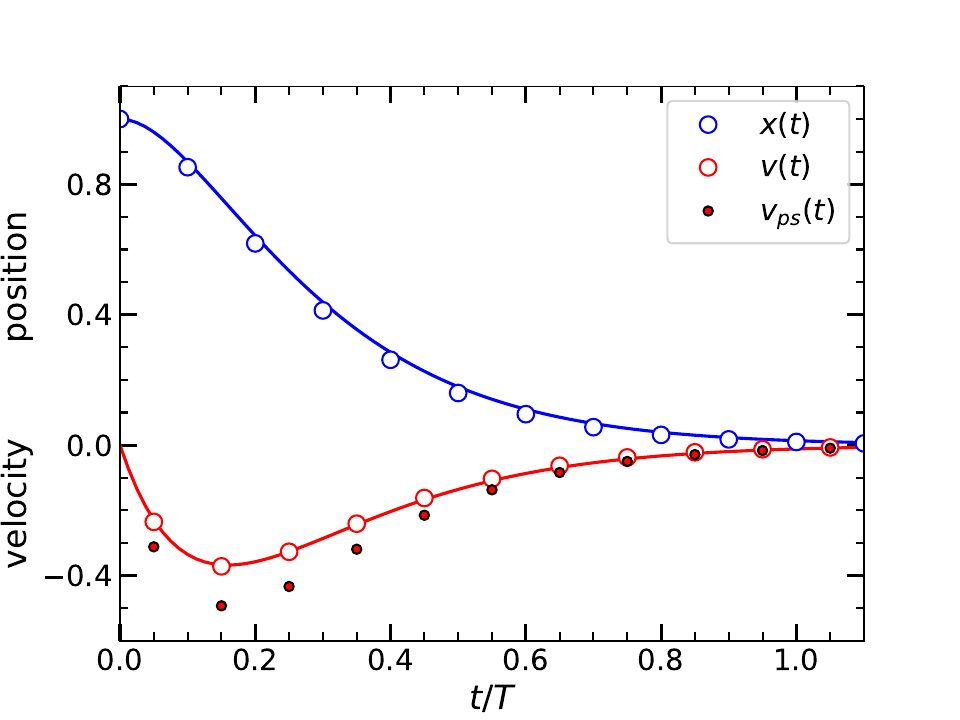}
\end{center}
\caption{ \label{fig:crit_ho_determ}
Exact solution (lines) of the critically damped harmonic oscillator with the initial condition $x(0) = 1$ and $v(0) = 0$.
Positions are shown on full time steps in blue and velocities on half time steps in red.
Data for numerical integration are based on the propagation coefficients of Grønbech-Jensen (circles). 
The pseudo-velocities are shown as small, full circles, the centered-differences of $x(t)$ as open, red circles.
}
\end{figure}

Using the just-described initialization scheme in the absence of thermal fluctuations produces the exact same $x_n$ sequence in the GJ integration scheme as in BPCL for a harmonic system.  
This can be readily verified by formally performing the respective steps by hand or, more effectively, by letting the computer do the work.  
Fig.~\ref{fig:crit_ho_determ} confirms this expectation for a critically damped harmonic oscillator with $\tau = \sqrt{k/m}/2$ and a time step of $\Delta t = T/10$, where $T = 2\pi\sqrt{m/k}$ is the period; when $T$ denotes temperature, it is preceded by $k_B$ or explicitly called temperature.

\subsection{Gaussian versus uniform random numbers}

\begin{figure}[h!]
\begin{center}
\includegraphics[width=0.45\textwidth]{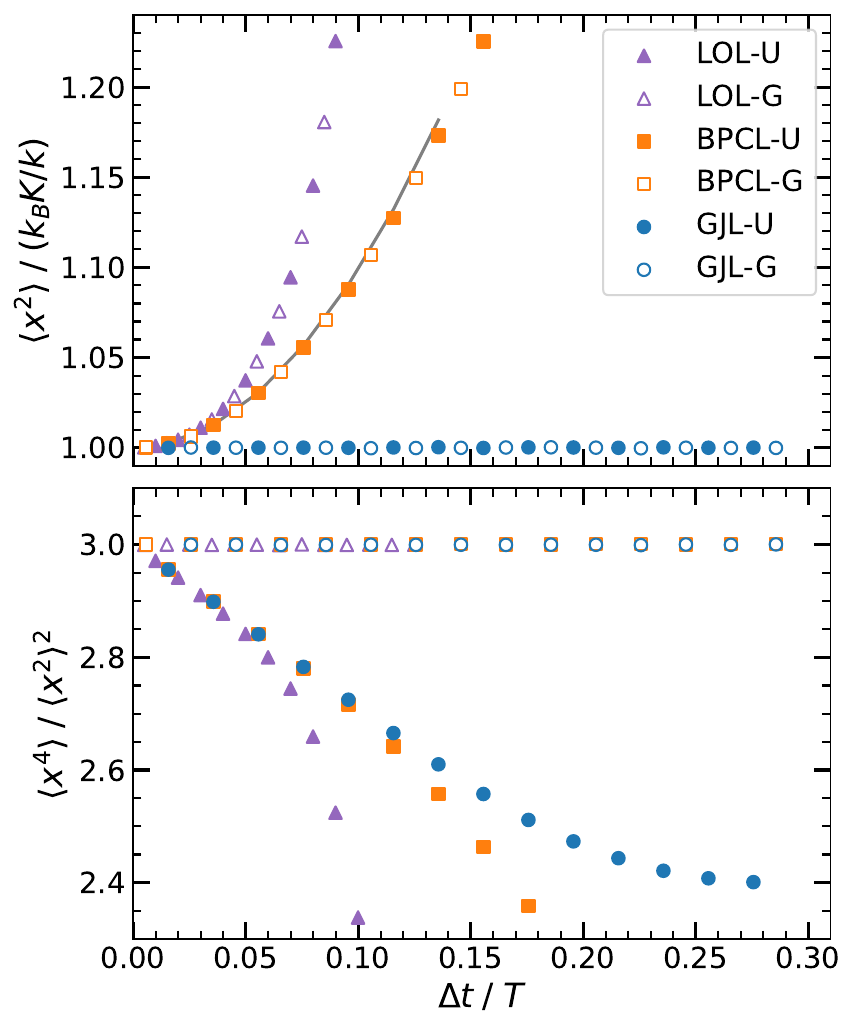}
\end{center}
\caption{ \label{fig:moments_ho}
Dimensionless second (top) and fourth (bottom) moments of a critically damped harmonic oscillator's coordinate $x$ as a function of the dimensionless time step $\Delta t / T$ using different thermostats.  
Grønbech-Jensen (GJL, blue circles), best-possible conventional (BPCL, orange squares), and lowest-order Langevin (LOL, purple trianles) thermostats, each time with Gaussian (-G, open symbols) and uniform (-U, full symbols) random numbers.
%
}
\end{figure}

It is occasional practice to use uniform random numbers on $[-\sqrt{3}, \sqrt{3}]$ in Langevin thermostats instead of regular Gaussian random numbers.  
Fig.~\ref{fig:moments_ho} examines the effect of this choice on the second and fourth moments of the position of the harmonic oscillator, from which its mean potential energy and specific heat can be deduced.  
As expected, the second moments are correctly reproduced in the GJ thermostats, irrespective of the higher moments of the random-number distribution, provided that its first moment vanishes and the second moment is unity.  
However, the fourth moment of the position is only correctly reproduced when the random numbers also have the correct fourth moment, which becomes evident when considering how to prove the correctness of the GJ approach through a moment analysis.  
Given that drawing random numbers constitutes a negligible fraction of the computational cost in the overwhelming majority of MD simulations, we highly recommend the use of Gaussian rather than uniform random numbers in conjunction with the GJ thermostat.  
Of course, ``exactness'' cannot be claimed based on numerical data alone.  
However, the numerical data deviates from exact values only within statistical errors;  
the number of correctly reproduced digits increases by one each time the statistics are increased by a factor of one hundred.  
This claim also holds for the sixth and higher moments, which is all that is needed from a practical perspective to validate that GJ indeed produces the exact harmonic distribution.  

Fig.~\ref{fig:moments_ho} also includes data for the BPCL and LOL thermostat.
Both conventional approaches quickly deviate from the exact expectation values due to the small time constant associated with critical damping, $\tau = T/(4\pi) \approx 0.08$.  
Differences between the two are marginal up to the point where errors in energy of 2\% might no longer be deemed acceptable.  
In this case, the fourth moment of the position $x$ is reproduced more accurately with uniform than with Gaussian random numbers, actually making uniform random numbers potentially the better choice for these two conventional Langevin-equation solvers owing to fortuitous error cancellation, at least for systems very close to being harmonic. 

\subsection{Convergence in anharmonic systems}

Although condensed-matter systems can usually be considered near-harmonic in some sense, most processes of academic interest, and all processes of practical relevance, involve anharmonic phenomena, such as local or collective transitions over energy barriers or thermal expansion.
Thus, good convergence in anharmonic systems is desirable, and the question arises as to how errors in static and dynamic properties depend on the time step size $\Delta t$ or the thermostat coupling constant $\tau$. 
We begin by discussing static properties, focusing on the potential energy, as this is the decisive quantity appearing in the Boltzmann factor.
To this end, we considered two systems: one in which convexity and potential energy are bounded and another in which both properties are unbounded. 
Specifically, we examined a single sinusoidal potential and Lennard-Jonesium in the second.
In the case of Lennard-Jonesium, we first study a linear chain of eight atoms and a bulk liquid next.
A liquid metal is also considered. 
The motivation for these choices is discussed where the models are introduced.\\

\noindent\textbf{Single-sinusoidal potential}\nopagebreak\vspace*{2mm}\par\nopagebreak
The primary reason why we study a single sinusoidal potential is that the convexity of the potential decreases as atoms move out of their equilibrium sites, while it increases for Lennard-Jonesium.
This allowed us to explore whether the sign of the leading errors in the GJ algorithm is sensitive to this property.
Moreover, the absolute value of the convexity is bounded, which can become problematic in the context of the LJ chain, as discussed further below.
This facilitates a rigorous convergence analysis of the studied algorithms in the single-sinusoidal potential.

The single-sinusoidal potential is given by
\begin{equation}
\label{eq:substrate_potential}
U(x) = -U_0 \cos(qx),
\end{equation}
which is a low-order but generic model to represent the interaction between adsorbed atoms and a crystal surface.
It is also a caricature model for the three-fold rotational barrier in polymeric backbones, such as the trans-gauche transitions in alkane chains or the rotation of methyl groups in side chains.
The maximum (absolute value of the)  convexity of the potential is $\kappa = q^2 \vert V_0 \vert$, which is attained at the extrema of the potential.
Its mean potential energy is easily shown to obey  
\begin{equation}  
U_\text{c}(T) = -U_0 I_1(V_0/k_B T) / I_0(V_0/k_B T),  
\end{equation}  
where $ I_n(x) $ is the modified Bessel function of the first kind of order $ n $.  

The thermostats are tested at a thermal energy of  $k_BT = 0.4~U_0$, which is very close to the point, where the specific heat assumes its maximum. 
%
At the investigated temperature, the mean potential energy is $U(T=0.4) = -0.765~U_0$.
%
%
Thus, the excess mean potential energy, $\Delta U = U(T)=U(0)$ divides into a harmonic and an anharmonic contribution of $0.2~U_0$ and $0.035~U_0$, respectively.
This division of close to 10\% energy corrections to a harmonic reference is not atypical for condensed matter near the melting point.
For example, both aluminum~\cite{Grabowski2009PRB} and Lennard-Jonesium~\cite{Della1998PRB} exhibit anharmonic energy corrections that are somewhat smaller than, but still close to, 10\% relative to the harmonic contribution.
Moreover, a ratio of $k_BT/U_0 = 0.4$ is roughly twice and thus of order the ratio of thermal energy and trans/gauche energy barriers. 
Thus, the error analysis is not merely a mathematical exercise but should be typical for some real systems.

\begin{figure}[h!]
\begin{center}
\includegraphics[width=0.45\textwidth]{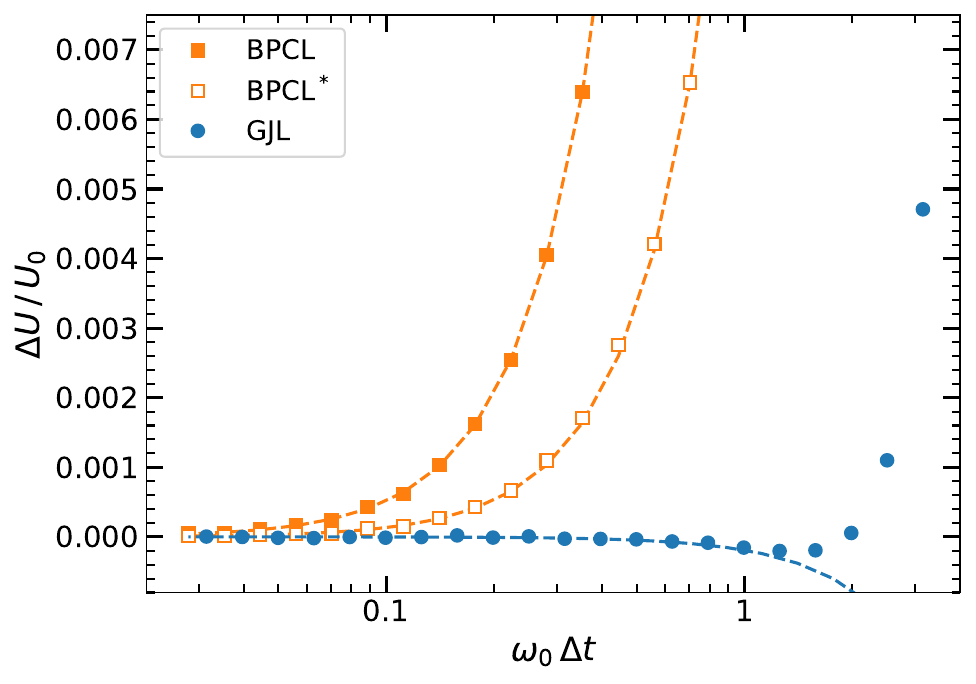}
\end{center}
\caption{ \label{fig:energy_substrate} Error in the mean potential energy $\Delta U = U_\text{MD}-U_\text{exact}$ at $k_BT = 0.4~U_0$ as a function of time step $\Delta t$ for the BPCL with and without correction, the latter being marked by a star and open squares, and GJL (circles).
The unit of time is the inverse of the circular frequency of the harmonic reference, $\omega_0^{-1} = \sqrt{m/V''_\text{max}}$.  
Dashed lines indicate the approximate leading-order error in the limit of small $\Delta t$, which are found to scale as $\Delta t^2$ in each case. 
}
\end{figure}

Fig.~\ref{fig:energy_substrate} shows errors in the mean potential energy for the BPCL and the GJ algorithm. 
The estimates for BPCL are augmented with second guesses for $U(T = 0.4)$, leveraging the system's near-harmonic behavior and the fact that, in the harmonic approximation, BPCL overestimates both potential and kinetic energies by the same amount.

This leads to a corrected estimate of 
\begin{equation}
\label{eq:harmonic_correction}
U(\text{BPCL}^*) = U(\text{BPCL}) +  \frac{k_BT}{2} - \langle T_\text{kin} \rangle.
\end{equation}
for the mean potential energy per propagated degree of freedom.
In principle, the correction could also have been multiplicative, $k_B T / (2 T_\text{kin})$, as this would be meaningful for two-point correlation functions, referring to two points in time or space.
However, an additive correction was chosen as an accurate reference energy is usually difficult to define, unless the system has a well-defined reference structure like a crystal. 

Various results in Fig.~\ref{fig:energy_substrate} are worth discussing.
First, all leading-order errors scale with the square of the time step size.
%
Second, incorporating the correction proposed in Eq.~\eqref{eq:harmonic_correction} reduces the leading-order error by close-to a factor of four.
Thus, for a given desired accuracy for the internal configuration energy, $\sigma_U$, the time step size $\Delta t$ can be increased by a factor of 2. 
Third, the leading-order error reduces by another factor slightly exceeding 80 when using GJ compared to the corrected BPCL results.
This non-trivial results means that $\Delta t$ can be increased by another factor of 9, if the desired target accuracy happened to be extremely small.
However,  for regularly used $\sigma_U$, corrections to the leading-order errors may no longer be negligible.
In fact, a property of the GJ scheme appears to be that there is a very thin domain, where $\Delta t$ leads to ``intermediate'' errors.
This behavior is borne out even more clearly for potentials with unbounded convexity.

We conclude the analysis of the single sinusoidal potential with some typical numerical numbers.
Fig.~\ref{fig:energy_substrate} is suggestive of an upper bound for the time step of 5\% of the period $T$, assuming that the error in $U(T)$ should remain below $0.005~k_BT$, which is 1\% of the mean potential energy relative to a minimum in the harmonic approximation. 
This translates to a BPCL time step of 1.5~fs for $T \approx 30$~fs, a typical value for a C-C bond vibration.
The time step could be doubled for BPCL* and further increased to slightly more than 15~fs when using GJ without deteriorating results, at which point the higher-order corrections to the exact potential energy are no more negligible.
However, this result certainly hinges on the anharmonicity.
In most cases, one should stay clear of the stability threshold for a harmonic vibration, i.e., not use $\Delta t > T_\text{min}/5$, when $T_\text{min}$ is the minimum (harmonic) period in the system. \\

\noindent\textbf{One-dimensional Lennard-Jonesium chain}\nopagebreak\vspace*{2mm}\par\nopagebreak
The curvature of the potentials considered so far is equal to or bounded by the value it takes in their minimum.
However, most MD simulations use potentials with unbounded convexity, such as Lennard-Jones.
%
Mean values are then no longer strictly defined, because there is a finite probability that two atoms assume identical Cartesian coordinates within data precision.
Practically speaking, such a freak event never happens, because its probability to occur is astronomically small.
This and more relevant but less serious problems arising due to large convexity can be averted, e.g., by modifying the repulsive part of the potential below a finite threshold distance. 

The mentioned issue arises even in a one-dimensional, on-site quartic oscillator, $V(x) = k_4x^4/4$:
for any finite time step, an atom reaching the $\Delta t$ dependent threshold $x_\text{t}$ at zero velocity will bounce back to a coordinate less than $-x_\text{t}$.
The condition for this and a subsequent built-up of amplitude to happen for large $\tau$ is approximately $2x_\text{t} = c_{xv} c_{vf} k_4 x_\text{t}^3$ so that $x_\text{t}^2 \approx 2m/(k_4\Delta t^2)$ for a large (Langevin) time constant $\tau$.
Thus, demanding $\exp(-\beta U(x_\text{t}))$ to be extremely small translates to $\exp(-\beta m / (k_4\Delta t^4)$ having to be less than, say, $10^{-81}$. 
This value would imply the constraint $\Delta t < \sqrt[4]{\beta m^2/k_4}/3$, which is readily achieved. 

In true equilibrium, events with the required small probability density do not noticeably affect averages, which is why their contributions can be ignored. 
Even events with a probability (density) of $10^{-10}$—which may well take place during the course of a simulation—will have only a marginal impact on averages.
Thus, rather than include time-extensive \texttt{if} statements in the code, we simply checked whether two LJ atoms in a chain swapped places during the simulation. 
Had this occurred, the time step would have been deemed too large to consider the result.  
In practice, it is hard to trigger the warning, as the zone in which atoms swap places while the simulation remains stable (without producing non-numerical values) is extremely narrow.

Whenever time steps were so large that simulations had blown up, results were simply ignored. 
This happens at smaller time steps when using GKECs than local damping schemes, since GKECs barely counteract the local build up of singularities.
This effect appears to dominate the possibility of extreme Gaussian random number in stochastic thermostats pushing adjacent atoms toward their singularity. 

Besides LJ atoms frequently being on a direct collision course with neighbors in a linear chain, one-dimensional systems quickly introduce sampling challenges. 
Long-wavelength modes, which couple only weakly to the remaining degrees of freedom, carry more weight in one dimension than in higher-dimensional space. 
As a result, weaknesses of thermostats are more readily exposed.

In our one-dimensional chain, only nearest neighbors interact via the Lennard-Jones potential
\begin{equation}
U(r) = U_0 \left\{ \left( \frac{r_0}{r}\right)^{12}  - 2 \left(\frac{r_0}{r}\right)^{6}\right\}.
\end{equation}
The curvature in the minimum of the potential is readily obtained as $U''(r_0) = U_0(12\cdot 13 - 2 \cdot 6\cdot 7)/r_0^2$ in the $m-n$ notation of the LJ potential.
This yields a spring constant of $k = 72 U_0/r_0^2$ so that the maximum frequency in the chain is $\omega_0 =  2 \sqrt{k/m} \to 12\sqrt{2U_0/m}/r_0$ in the harmonic approximation.
This is due to the dispersion relation of a linear chain, $\omega^2(q) = 4(k/m) \sin^2(qa/2)$, where $q$ is the wavenumber and $a$ the nearest-neighbor spacing. 

According to the Lindemann criterion, a crystal melts when (relative) rms atomic displacements are about 10\% from its lattice site.
In an Einstein-solid approximation of our chain, these amplitudes are reached near $k_BT_\text{m} \approx 0.7~U_0$.
%
%
This is a plausible value for argon, which melts at about $70$~K and is frequently modeled using $U_0 \approx 120$~k$_B$K.
As a compromise between these two values, $k_B T = 2 U_0 / 3$ is used in the following.
%
To ensure that increasing convexity is sampled, thermal expansion is ignored and the lattice constant confined at $a = r_0$.

\begin{figure}[h!]
\begin{center}
\includegraphics[width=0.45\textwidth]{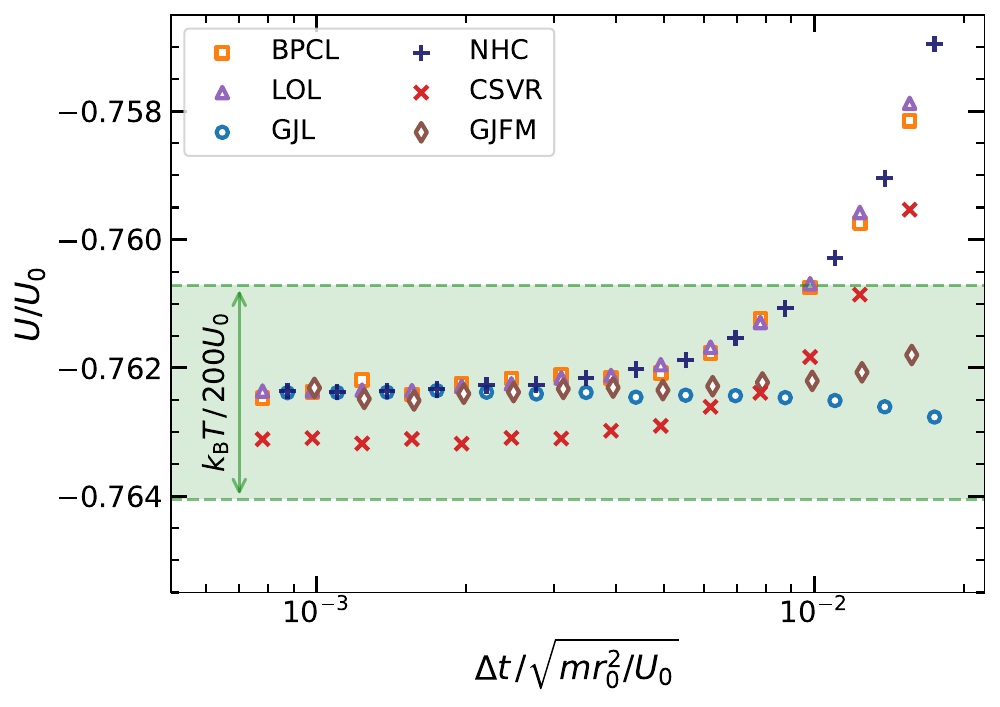}
\end{center}
\caption{ \label{fig:energy_lj_chain}
Mean potential energy $U$ of a Lennard-Jones chain
at $k_BT = 2 U_0 / 3$ %
as a function of time-step size $\Delta t$ for a variety of thermostats and a fixed thermostat time constant of $\tau = \sqrt{mr_0^2/U_0}/3$.
The Nosé-Hoover chains contained $m = 3$ elements. 
%
The center of the green band was obtained by fitting a quadratic function to the GJL data.
Its width is set to $k_\mathrm{B}T/(200U_0)$, such that the maximum relative error is 0.25\% of the excess configurational energy of a harmonic reference.
}
\end{figure}

Fig.~\ref{fig:energy_lj_chain} presents results on the dependence of the mean potential energy on time step size for a variety of thermostats.
They were obtained using a thermostat time constant $\tau = \sqrt{mr_0^2/U_0}/3$. Moreover, for GJFM a Maxwell spring of stiffness $k_\text{Mxw} = k/8$ was employed.
The results for CSVR, GJL, LOL and NHC were obtained using LAMMPS, while BPCL and GJFM were generated with our house-written Python code, which yielded statistically identical results for LOL and GJ as LAMMPS.
%
%
%
%
%
All thermostats seem to converge to the exact value for small $\Delta t $, except for CSVR, which produces an error close to 0.1\% of the thermal energy in the $\Delta t \to 0$ limit. 

A few more results contained in Fig.~\ref{fig:energy_lj_chain} are worth noticing. 
NHC and the two conventional thermostats LOL and BPCL reveal similar leading-order errors. 
This might not be surprising given that the thermostat time constant $\tau$ is large compared to the periods of local vibrations so that the perturbation to a Verlet or symplectic Euler scheme is small. 
GJ and likewise GJFM is again far more accurate than the other approaches. 
Demanding the systematic error in the mean potential energy to be 0.0025~$k_BT$ requires the time step to remain below $0.01\,\sqrt{mr_0^2/U_0}$ for NHC and conventional Langevin algorithms.
The two GJ-based schemed are still stable at twice that time step but suffer from higher-order errors at larger time steps. 
%
%
%
The gain is understandably not quite as large as for the single-sinusoidal potential.
Yet, a time step twice that of what is usually considered an aggressive choice still gives highly accurate mean potential energies, roughly ten times more accurate than for the other approaches. 

\begin{figure}[h!]
\begin{center}
\includegraphics[width=0.45\textwidth]{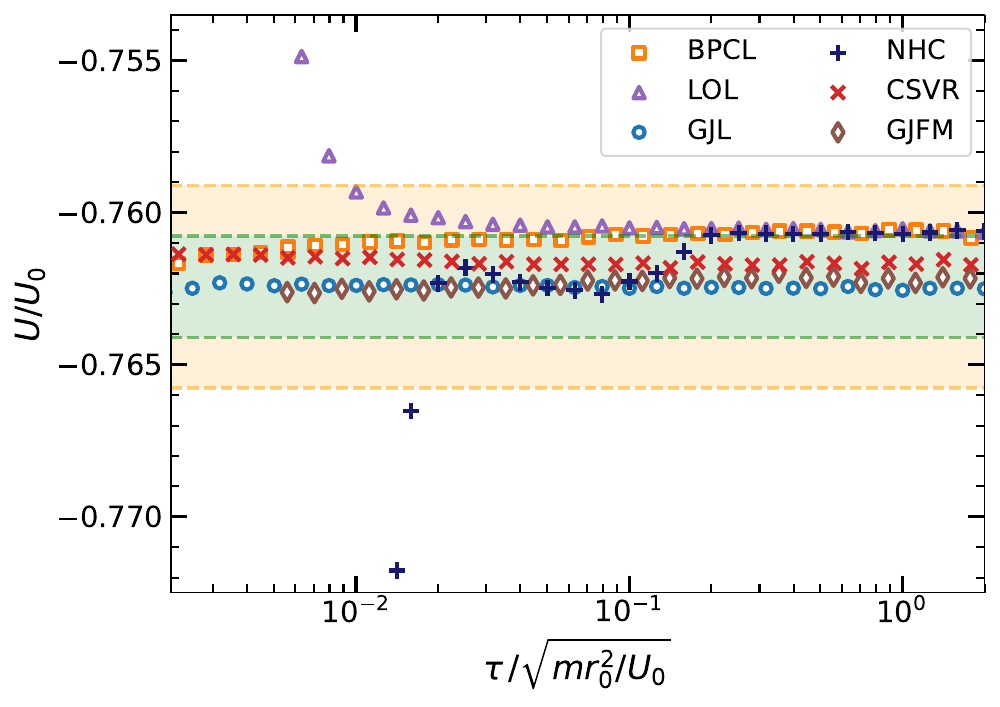}
\end{center}
\caption{ \label{fig:energy_lj_of_tau}
Similar to Fig.~\ref{fig:energy_lj_chain}, however, this time energy $U$ as a function of time constant $\tau$ at a fixed time step of $\Delta t = 0.01\sqrt{mr_0^2/U_0}$.
The green band is identical to that of Fig.~\ref{fig:energy_lj_chain}. 
The width of the orange band is twice that of the green one, and thus represents a deviation in the total potential energy of $\pm k_\mathrm{B}T\,/\,(200\, U_0)$.
}
\end{figure}

Fig.~\ref{fig:energy_lj_of_tau} explores the dependence of the mean potential energy on the thermostat time constant $\tau$ at a fixed time step of $\Delta t = 0.01~\sqrt{m r_0^2/U_0}$.
CSVR, which benefits from fortuitous error cancellation at this large value of $\Delta t$ is well-behaved as $\tau$ tends to zero, i.e., errors do not start to diverge as they do for LOL and NHC.
The deviation from producing stable means is somewhat unusual for NHC, as the mean potential energy first transitions from one stable to another stable value upon decreasing $\tau$ before it starts diverging. 
BPCL starts to approach the supposedly exact value at the smallest investigated $\tau$.
However, it is clear that BPCL, just like GJ, will become exact for $\tau \to 0$, since they both transition to Brownian dynamics with fxed $\Delta t$ but increased damping. 
For GJ, this approach is difficult to demonstrate, as systematic errors are too small to be resolved within reasonable simulation times. 
%
%
%
\\

\noindent\textbf{Bulk liquid Lennard-Jonesium}\\

We next consider liquid Lennard-Jonesium in three spatial dimensions and explore to what degree previous results extend to bulk systems.
To this end, a cutoff radius of $r_\text{c} = 2.5~\sigma_\text{LJ}$ is used.
The thermal energy remains at $k_B T = 2U_0/3$.
The simulation cell is cubic.
It contains $N = 1,000$ atoms.
Its edge length is $10.817~\sigma$, at which point the mean pressure is a small fraction of $U_0/r_0^3$ at the given temperature.

\begin{figure}[hbtp]
    \includegraphics[width=0.45\textwidth]{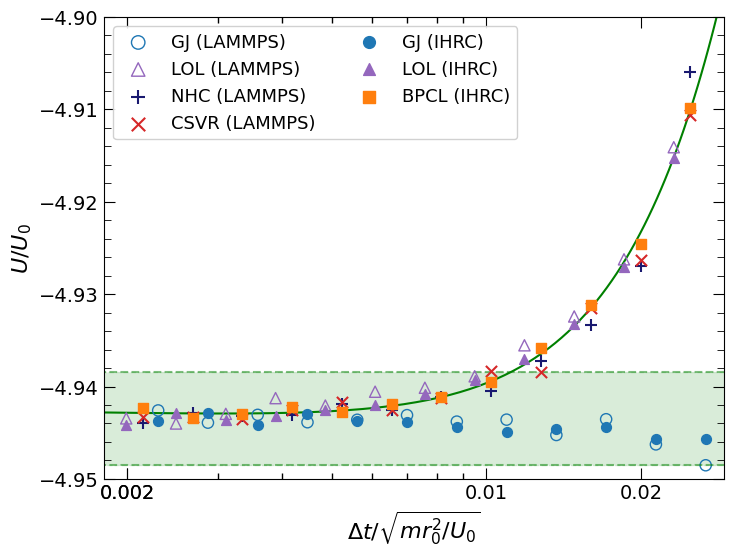}
    \caption{\label{fig:bulk_lj_dt}
    Mean potential energy $U$ as a function of the timestep $\Delta t$ for different thermostats and a fixed thermostat time constant $\tau=0.5$. Solid markers represent measurements taken with our house-written Rust code. The solid line represents quadratic fit to the non-GJ thermostats to better guide the eye, while the green area represents the acceptable deviation range $\Delta U = \pm  D k_B T / 400 $ per atom, where $D=3$ is the spatial dimension.
    }
\end{figure}

Results for three-dimensional Lennard Jonesium are shown in Fig.~\ref{fig:bulk_lj_dt}.
They were produced using LAMMPS for GJ, NHC, and CSVR, and with a house-written Rust code for BPCL.
The latter also has an implementation for GJ, which, produces statistically identical results as LAMMPS.
In addition, we consider the \textit{regular Langevin} (RL) thermostat implementation in LAMMPS, which, however should be isomorphic to our LOL approach.

The results for three-dimensional systems corroborate those found in $D = 1$.
However, relative errors are much larger now, so that conclusions cannot be drawn with the same rigor as for the one-dimensional LJ chain. 
Moreover, the relative importance of long-wavelength, quasi-harmonic modes, which GKECs do not sample properly, is less in higher dimension.
This might explain why the edge that GJ has on more established thermostat appears to be smaller in three-spatial dimensions than in one.\\

\noindent\textbf{Bulk liquid metal}\\

\begin{figure}[hbtp]
    \includegraphics[width=0.45\textwidth]{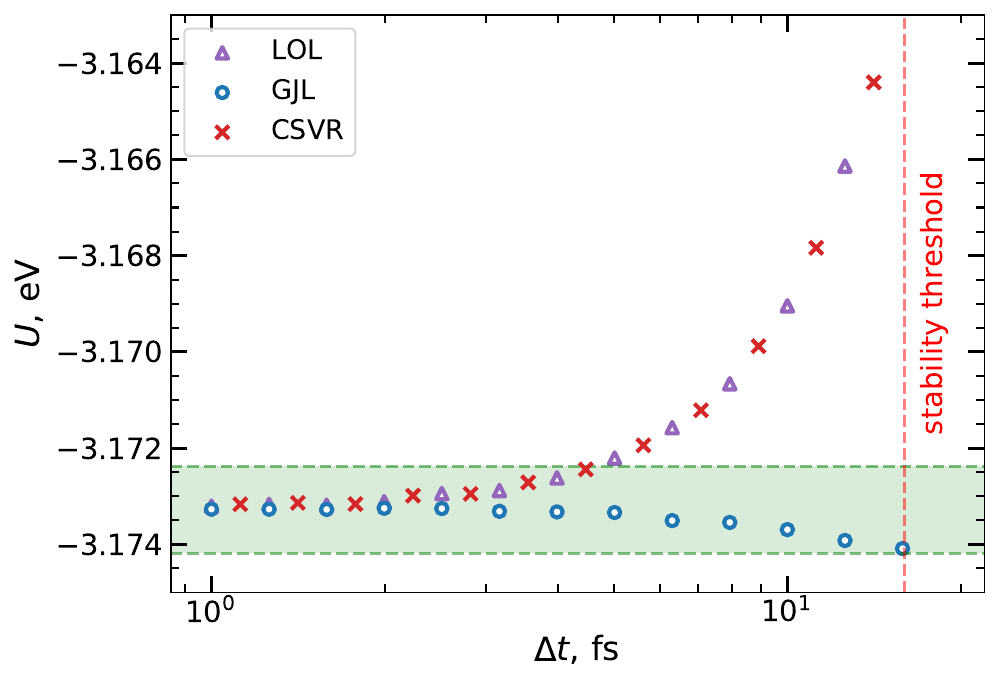}\caption{\label{fig:copper}
Mean potential energy $U$ per copper atom as a function of the time step $\Delta t$ for three selected thermostats. A fixed value for the thermostat time constant was used $\tau = 6\tau_\mathrm{D}$, with $\tau_\mathrm{D} = 78.74$~fs being the Debye time of copper. The temperature was chosen as $T=1400$~K. The green area represents the acceptable deviation range of $\Delta U = \pm Dk_\mathrm{B}T/400$ with $D=3$.
}
\end{figure}

The repulsive interaction between LJ atoms is too stiff compared to realistic potentials, which assume a more exponential dependence at typical particle spacings. 
To explore the benefits of advanced over traditional thermostats for more realistic repulsion and to extend our analysis from soft to hard matter, we also simulate a liquid metal above its melting temperature.
Specifically, we use a generic second-order tight-binding potential, which for a mono-atomic system is equivalent to the embedded-atom method, and use a parameterization adopted to the simulation of copper~\cite{Jalkanen2015MSMSE}.  
Fig.~\ref{fig:copper} shows simulation results obtained at a temperature of $T = 1,400$~K for $N = 4,000$ atoms and using a particle density of $\rho = $74.1~nm$^{-3}$.
For these choices, the mean pressure is well below 50~MPa.

Fig.~\ref{fig:copper} reveals that time steps of 5--10~fs are readily possible when simulating copper using GJL, while most studies use time steps from 1--2~fs. 
An interesting feature of GJL is that the method either produces highly accurate results, i.e., with errors in the potential energy being at most $0.0025~k_BT$ per degree of freedom, or it is unstable.
This means that results, given they are produced, are reliable.
As was the case for bulk LJ liquids, CSVR and LOL are again on par.
They both become unstable at the same time-step size, where GJL becomes unstable.
However, if larger damping is used, the stability range of GJL (and likewise of BPCL) increases while that of LOL would decrease. 

We note in passing that the amount of anharmonicity in liquid copper was evaluated to be 8\% of the harmonic contribution in the liquid phase just above the melting temperature.
This is on par with the ratio reported for aluminum in the solid phase just below the melting point in the crystalline phase~\cite{Grabowski2009PRB}.
In our case, the number was obtained by quenching 3{,}000 independent liquid samples to the closest energy minimum using conjugate-gradient minimization.
The potential energy of the basin of attraction was subtracted from the mean potential energy of the liquid, which then allowed the anharmonic contribution to be deduced. 
The rather minor degree of anharmonicity in a typical liquid may explain the excellent results obtained with the GJ algorithm.

\subsection{Mean-squared displacements}

\begin{figure}[hbtp]
    \includegraphics[width=0.45\textwidth]{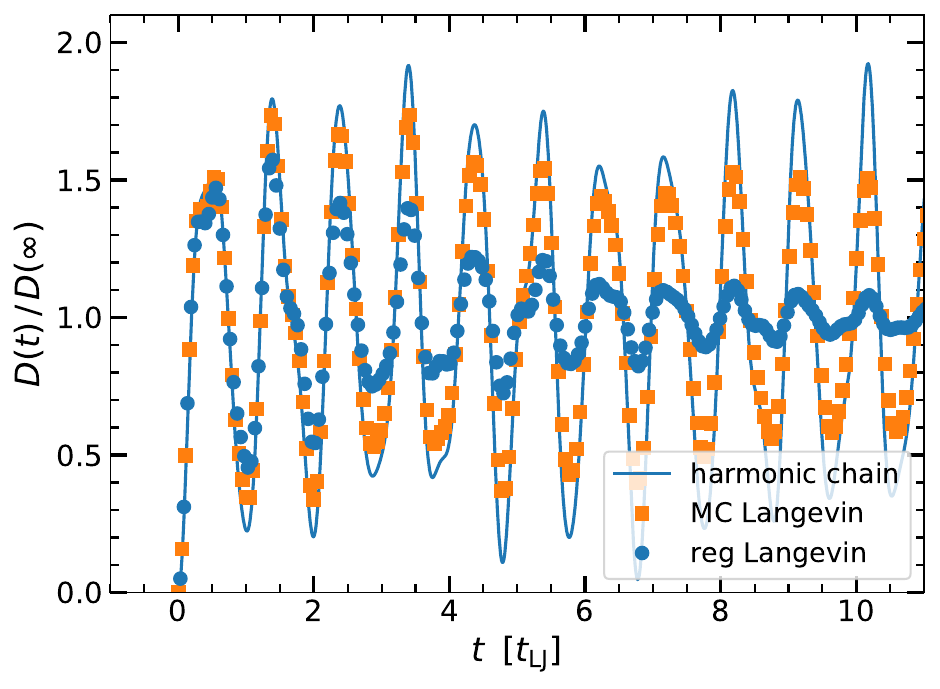}
    \caption{\label{fig:diffused_dist}
    Mean-square displacement $D(t)$ as a function of time $t$ of a linear Lennard-Jones chain consisting of $N = 8$ atoms at a thermal energy of $k_B T = 0.001~\varepsilon_\text{LJ}$.
    Data for the momentem-conserving (MC) Langevin thermostat are shown in orange square, for a regular (reg) Langevin by purple circles.
    The analytical solution of a periodically repeated, linear-harmonic chain is included for comparison.
    }
\end{figure}

As mentioned several times in this work, one significant drawback of regular Langevin thermostats is that they act in a frame of reference, whereby long-wavelength vibrations become overdamped.
The momentum-conserving Langevin thermostat introduced here mitigates this problem dramatically, as is demonstrated in Fig.~\ref{fig:diffused_dist}.
It shows the (relative) mean-square displacement of atoms in the chain
\begin{equation}
D(t) = \langle \left\{ x(t) - x(0)\right\}^2\rangle,
\end{equation}
where $x(t)$ is an atom's position relative to the center of mass of the chain.
Both the regular and the momentum-conserving (MC) thermostats use the same damping time, which is set to be twice the largest time in the system.
This means that \emph{all} modes are underdamped; however, those propagated by the momentum-conserving Langevin thermostat are significantly less damped than those of the regular one, since the quality factor of the new thermostat increases with increasing wavelength.
The effect is quite noticeable, although the chain is relatively short with $N = 8$ atoms.

\section{Conclusions}
\label{sec:conclusions}

In this work, we revisited, extended, and evaluated various thermostatting schemes, primarily those based on the Langevin equation. 
Special emphasis was placed on implementation, analytical grounding (i.e., asymptotic analysis instead of short-time propagator decomposition), and the need to thermostat quasi-harmonic modes, which couple only weakly to the remaining degrees of freedom.
Among conventional Langevin thermostats—i.e., those that do not require random numbers to be retained between time steps—we identified the best-possible conventional Langevin (BPCL) scheme.
Like the Grønbech-Jensen (GJ) algorithm, BPCL seamlessly converges to Brownian dynamics as the mass $m$ tends to zero while the damping $\gamma = m/\tau$ remains constant, and it does so in a way that permits an increase in time step during this transition, as the dynamics become isomorphic to Brownian motion.
This property is difficult to achieve with schemes in which $m$ appears in the denominator of a propagation coefficient, or where velocities are \emph{true} velocities, whose thermal fluctuations diverge for small $m$.
However, at small damping, e.g., when the thermostat time constant clearly exceeds the Debye time of a solid, BPCL offers no advantage over LOL. 

The GJ algorithm demonstrated superior performance for harmonic and near-harmonic systems, yielding correct thermal averages across a broad range of damping constants and time steps. 
While it slightly complicates the update rules by recycling random numbers, it does so to great benefit.
GJ's high accuracy persists even in moderately anharmonic systems, such as a sinusoidal potential or liquid Lennard-Jonesium, where it allows time steps that exceed conventional stability limits by factors of two to ten. 
Another example is liquid copper at 1{,}400~K and ambient pressure, where a time step of 10~fs still yields highly accurate thermal averages.
A highly beneficial feature of GJ-based thermostats, when applied to realistic potentials, is that there exists only a very small — or even non-existent — range of time steps for which the algorithm remains stable but produces inaccurate averages. This implies that if the simulation runs stably, one can generally assume that time-step discretization errors are small.

One promising avenue is to apply a predecessor of GJ—namely the Brownian thermostat by Grønbech-Jensen and Farago (GJF)—to Maxwell elements, which then act as thermostats for the true degrees of freedom.
This approach combines the precision of GJ with the smoother trajectories typical of deterministic solvers, thereby avoiding the short-time erratic behavior of conventional Langevin methods.
The latter can be inconvenient when extended Lagrangian techniques are used.

We also proposed a lean momentum-conserving Langevin (MCL) thermostat, which thermostats atoms relative to the center of mass in small spatial bins. 
This scheme preserves long-wavelength vibrational modes and hydrodynamic interactions, avoiding the overdamping artifacts introduced by thermostats acting in a fixed frame of reference. 
MCL is even simpler to implement than DPD and generally requires fewer floating-point operations.
It can also be extended in various directions.
For example, MCL can be made hierarchical: the centers of mass may be thermostatted with BPCL rather than Verlet integration, using masses chosen such that characteristic frequencies at different length scales collapse as closely as possible.
A convenient choice for fast equilibration is to set the center-of-mass inertia proportional to the (hyper)surface area of the bin at the given scale.
Another natural extension of the MCL approach is its application to molecules or covalently bonded sub-units, such as all-atom CH$_3$ groups or flexible H$_2$O models.
The undesired excess viscosity and reduced diffusion coefficient—arising from the lack of angular-momentum conservation—could be offset by assigning smaller masses to the thermostatted centers.

In summary, thermostat choice should be guided by the physical properties of interest.
When true equilibrium dynamics (in undriven systems) are of concern, the safest option is to use a \emph{real} thermostat—rather than a global kinetic energy control—and to average over different initial conditions.
If this is not appealing and natural dynamics at large or mesoscopic scales are of interest, we find that a momentum-conserving Langevin-based approach is best.
Another viable option is to combine lowest-order Langevin (LOL) thermostats: one acting in the frame of reference with a small time constant chosen to critically damp long-wavelength modes, and another momentum-conserving one (MCL or DPD in LOL form) with a much shorter time constant to quickly equilibrate local modes.

\section*{AUTHOR DECLARATIONS}

\subsection*{Conflict of Interest}
The authors have no conflicts to disclose.

\subsection*{Author Contributions}
MHM conceptualized the study and led the development of the manuscript.
All authors contributed to the research through code development, simulations, data analysis, and figure preparation.
SA and SVS also contributed to manuscript revision.
All authors discussed the results and approved the final version of the manuscript.

\subsection*{Data Availability}
The in-house codes, LAMMPS input files, and the data that support the findings of this study are available from 
\url{https://github.com/mueser/Adv_Lang_Thermostats.git}

\bibliography{langevin}

\end{document}